\documentclass[a4paper,pra,showpacs,nofootinbib,onecolumn]{quantumarticle}

\pdfoutput=1  %Suggested for quantumarticle

\usepackage{enumitem}
\usepackage{amsmath}
\usepackage{amssymb}
\usepackage{amsthm}
\usepackage{graphics}
\usepackage[pdftex]{graphicx} % Load graphicx with pdftex option
\usepackage{epstopdf} % Load epstopdf after graphicx
\usepackage{subcaption}
\usepackage{color}
\usepackage{mathrsfs}

\usepackage[usenames,dvipsnames,svgnames,table]{xcolor}
\usepackage{stmaryrd}
\usepackage{calc}
\usepackage{url}
\usepackage{hyperref}
\long\def\ca#1\cb{}

\def\bra#1{\langle#1|}
\def\dyad#1#2{|#1\rangle\langle#2|}
\def\inpd#1#2{\langle#1|#2\rangle }
\def\ket#1{|#1\rangle }

\def\Tr#1{\textrm{Tr}\left(#1\right)}
\def\pTr#1#2{\textrm{Tr}_{#1}\left(#2\right)}

\def\locc{\overline{\textrm{LOCC}}}  % closure of LOCC

\def\myeq#1{Eq.~\eqref{#1}}
\def\AC{{\cal A}}
\def\BC{{\cal B}}

\def\EC{{\cal E}}

\def\FC{{\cal F}}

\def\HC{{\cal H}}
\def\IC{{\cal I}}
\def\JC{{\cal J}}

\def\MC{{\cal M}}

\def\PC{{\cal P}}

\def\endproof{{\hspace{\stretch{1}}$\blacksquare$}}

\newcommand{\norm}[1]{\left\vert#1\right\vert} % for ordinary vectors
\newcommand{\normm}[1]{\left\|#1\right\|} % for larger things

\newtheorem{thm1}{Theorem}

\newtheorem{thm2}[thm1]{Theorem}

\newtheorem{thm6}[thm1]{Theorem}

\newtheorem{thm8}[thm1]{Theorem}

\newtheorem{lem1}{Lemma}

\newtheorem{lem2}[lem1]{Lemma}
\newtheorem{lem3}[lem1]{Lemma}
\newtheorem{lem4}[lem1]{Lemma}
\newtheorem{lem5}[lem1]{Lemma}
\newtheorem{lem6}[lem1]{Lemma}

\newtheorem{lem9}[lem1]{Lemma}

\begin{document}
	\title{Asymptotic implementation of multipartite quantum channels and other quantum instruments using local operations and classical communication}
	\author{Scott M. Cohen}
	%\ca
	\email{cohensm52@gmail.com}
	\affiliation{Department of Physics, Portland State University, Portland Oregon, USA 97201}

	\begin{abstract}
		\noindent We prove a necessary condition that a quantum channel on a multipartite system may be approximated arbitrarily closely using local operations and classical communication (LOCC). We then extend those arguments to obtain a condition that applies to all quantum instruments, which range from the most refined case, a generalized measurement, to the most coarse-grained, the latter being a quantum channel. We illustrate these results by a detailed analysis of a quantum instrument that is known not to be implementable by LOCC, but which can be arbitrarily closely approximated within that framework. As one outgrowth of this analysis, we find a quantum measurement that falls into the same category: it cannot be implemented exactly by LOCC, but can be approximated by LOCC arbitrarily closely. This measurement has an infinite number of outcomes, pointing to the intriguing question as to whether or not there exists a measurement within this same category but having only a finite number of outcomes.
	\end{abstract}
	
	\date{December 5, 2025}
	\maketitle
	
	\section{Introduction}\label{sec1}

	In two recent papers \cite{myProdPaths,myProdPathsDiscriminate}, we developed a new approach to the study of quantum measurements implemented by multiple parties utilizing \emph{local operations} in their respective laboratories and \emph{classical communication} to their partners to share the results of those local operations, a fundamental paradigm in quantum information processing commonly known as LOCC. We believe our approach provides valuable insights and tools for analyzing what can be accomplished under these circumstances, where quantum systems are not (under practical conditions, often \emph{cannot}) be transmitted from one party to another. In the first of these papers \cite{myProdPaths}, we identified certain geometric conditions that must hold in order that a measurement can be approximated arbitrarily closely by LOCC, and then in the second paper \cite{myProdPathsDiscriminate}, we refined that approach for use in studying local quantum state discrimination, showing how one can often determine whether success with such a task is impossible, even while knowing only the set of states to be discriminated but nothing about which measurements, if any, might be possible candidates for successfully accomplishing such a task. Here, we generalize our approach to address the question of whether or not a given quantum channel can be closely approximated using LOCC. We then extend the ideas further to apply to any quantum instrument \cite{WinterLeung}.
	
	The paradigm of LOCC has been widely studied \cite{Bergou,BennettTele,Bennett9,BennettPurifyTele,Walgate,CiracDistComp,Nielsen,BKrausStateTransf,ChitambarHsiehHierarchy,ChildsLeung,KKB,FuLeungMancinska,ChitambarHsiehPeresWootters,myQChImpossbyLOCC,myStrongBounds,ChildsLeung} since publication of the seminal paper by Peres and Wootters \cite{PeresWootters}. While much progress has been made, and many important and sometimes surprising results have been obtained, LOCC has also been widely recognized as being quite difficult to characterize. One particularly challenging aspect arises when one is interested in whether or not a task can be accomplished while allowing an arbitrarily small error to be incurred in doing so. In studying such questions, one must allow the parties to share their outcomes back and forth for as many rounds of communication as needed in order to reduce the error as much as possible. With some tasks that can be easily accomplished by acting globally on the entire multipartite system, it is known that when restricted to LOCC, even allowing for an infinite number of rounds is not enough to reduce the error to zero. One such celebrated example involves perfect discrimination of the so-called domino states \cite{Bennett9}, which demonstrated that separable measurements---ones that involve measurement operators that are simple tensor products on a multipartite system and can be easily implemented globally---cannot always be implemented, or even closely approximated, by LOCC, even with the use of an infinite number of communication rounds. This discovery gave birth to a wide-ranging effort at understanding the surprisingly rich and difficult questions concerning the difference between global actions and those restricted to LOCC. 

	While LOCC has a structure that is difficult to characterize, one commonly discusses it in terms of a relatively simple picture, that of a rooted tree graph, or \emph{LOCC tree}. In the case of a finite LOCC tree and starting at the root node, one may follow one of the tree's many branches to its endpoint, which is known as a leaf node. Each node along any given branch represents an outcome of a (local) measurement performed by one of the parties, and the next node along the branch represents an outcome obtained in a subsequent (local) measurement by a different party. These nodes are connected by edges, and it is easy to envision a path following this branch along these successive edges from the root node all the way to its leaf. We recognized \cite{myProdPaths} that this path is not just an abstract visualization of the sequence of outcomes, otherwise having no useful meaning, but rather that it corresponds directly to a particular type of path lying within a specific geometric object. Significantly, that object is determined uniquely by the outcomes associated with the leaf nodes appearing at the ends of the collection of all branches in the tree. That is to say the measurement $\MC$ implemented by the LOCC protocol represented by this tree uniquely identifies a geometric object $Z_\MC$, within which lies each and every path associated with the branches in the tree. We have characterized these paths \cite{myProdPaths} and used that characterization to obtain a powerful condition, reproduced below as Theorem~\ref{thm2}, which must hold if it is possible for $\MC$ to be implemented arbitrarily closely by LOCC. When this condition is not satisfied, then, $\MC$ is not achievable within the closure of LOCC, which we denote by $\locc$. This approach has been illustrated in \cite{myProdPaths,myProdPathsDiscriminate}, where for numerous examples of local quantum state discrimination, it is shown that they are not achievable within $\locc$. This includes \textit{all} such cases that we were aware of at the time, including some that were previously not known to be outside $\locc$. In particular, we have there shown that all unextendible product bases discussed in \cite{IBM_CMP} are not distinguishable within $\locc$.
	
	Here, we extend that analysis beyond measurements to encompass the more general notion of quantum channels and indeed, any quantum instrument \cite{WinterLeung}. For channels, we find a condition, see Theorem~\ref{thm1} below, which must hold if quantum channel $\EC$ can be closely approximated by LOCC. The condition is a direct generalization of that for measurements in Theorem~\ref{thm2}. It also involves characterization of the paths discussed in the preceding paragraph, which in similar fashion must lie within a more general geometric object $Z_\EC$, which itself is uniquely determined by $\EC$. The difference in structures of $Z_\EC$ and $Z_\MC$ reflects the fact that in implementing a quantum channel, no classical information about outcomes can be available, in contrast to measurements, where complete (classical) knowledge of those outcomes must be known. Both channels and measurements are special cases of quantum instruments, which make allowance for the amount of available classical information to vary from all to none.
	
	We have found that dealing with $Z_\EC$ involves additional difficulties as compared to $Z_\MC$. As we shall see, the latter is a special case of the former, and it is this greater generality of $Z_\EC$ which makes it harder to analyze. Furthermore, whereas it was quite easy to find examples of previously studied measurements where we could apply, and learn from, our condition for measurements, Theorem~\ref{thm2}, examples of quantum channels to which we can fruitfully apply Theorem~\ref{thm1} are much more difficult to find. Perhaps the explanation for this dearth of examples is that quantum channels---in particular, multipartite quantum channels---are just more difficult to study, in general, so comparatively little work has been done in this direction. One may hope that the results presented here will stimulate further efforts in that direction.
	
	A dearth is not the same as a complete absence, however. In fact, there exists a two-qubit example of a quantum instrument for which it is known that it cannot be implemented exactly but can be approximated arbitrarily closely by LOCC \cite{WinterLeung}. The inspiration for this instrument came from ideas developed in a series of earlier papers on the random distillation of entanglement from a three-qubit system onto two of those qubits \cite{FortescueLo,Chitambar,ChitCuiLoPRA,ChitCuiLoPRL}. We study the two-qubit example in detail below, first as a quantum channel, then as a general instrument, and finally as a quantum measurement. This analysis leads to insights regarding how the critical property of our paths, that of being \emph{piecewise local}, see below, disappears in the limit as the number of rounds tends to infinity. It also illustrates other properties of these paths, such as the fact that certain matrices associated with these paths must have a block diagonal form determined by the structure of the given instrument in terms of its constituent completely positive (CP) maps. We also find from this example an interesting result regarding the closure of the set of LOCC measurements. Specifically, we will show that the set of LOCC measurements is not itself closed.
	
	The relationship between LOCC and its closure, $\locc$, has long been a point of interest to researchers. An early result along these lines was the demonstration, in the context of random distillation of two-party entanglement starting from the three-party $W$ state on qubits, that a convergent sequence of transformations is not achievable by LOCC in the limit \cite{ChitCuiLoPRL}. There, for a specific transformation, the authors provided a sequence of LOCC protocols that achieved that tranformation with probability arbitrarily close to $1$, but they also showed that this tranformation could not be achieved using LOCC with success probability exactly equal to $1$. It may not have been entirely clear to readers how to understand the limit of that sequence of protocols whose limiting protocol is not LOCC. So, yes, it fails to be an LOCC protocol, but is it a protocol at all? We demonstrate below that it is, indeed, a protocol, but one that is not amenable to local measurements; see Eqs.~(\ref{eqn3401}) and (\ref{eqn3402}) and the accompanying discussion, below.
	
	Later, utilizing a two-party generalization of those arguments, it was shown that LOCC is also not closed for two-qubit systems \cite{WinterLeung}, when the parties are attempting to implement a general quantum instrument. In each of these cases, the parties must rely upon classical information obtained in these protocols, leaving open the question of whether or not LOCC is closed when considering quantum channels, for which no classical information is retained. It has also been an open question whether LOCC is closed as applied to quantum measurements, where the maximum amount of classical information is needed to uniquely identify each and every measurement outcome. Here, in the context of the two-qubit example just mentioned, we discover a measurement which is not in LOCC but is within its closure. As far as we are aware, this is the first demonstration of a measurement exhibiting such a property.

	The remainder of the paper is organized as follows. In the next section, we provide an overview of the tools we will use in the subsequent sections, including a discussion of quantum channels and instruments and their mathematical representation, finite-round LOCC and its extension to infinite rounds, and the geometric objects, $Z_\MC$ and $Z_\EC$, mentioned above. In Section~\ref{sec3}, we present our main result, Theorem~\ref{thm1}. Section~\ref{sec4} provides a proof of Theorem~\ref{thm1}, which utilizes a result linking convergence of a sequence of quantum channels to convergence of a sequence of the geometric objects, $Z_\EC$. The latter result is proved in Appendix~\ref{secA}. In Section~\ref{sec5} the reader will find an extensive analysis, in the context of Theorem~\ref{thm1}, of the two-qubit example channel from Ref.~\cite{WinterLeung}. Section~\ref{sec9} provides a generalization of Theorem~\ref{thm1} to the case of arbitrary quantum instruments, encompassing and interpolating between Theorem~\ref{thm2} and Theorem~\ref{thm1}. There, the reader will also find a brief discussion in this context of the two-qubit example of Section~\ref{sec4}. The important question of whether or not LOCC is a closed set, a question that may depend on the type of circumstances to which it is being applied, is addressed in Section~\ref{sec6} where it is shown that LOCC is not closed when applied to quantum measurements. Section~\ref{sec7} presents a generalization of the example of Section~\ref{sec4} to any number of qubits, and then finally, we offer our conclusions in Section~\ref{sec8}.
	
	\section{Preliminaries}\label{sec2}
	In this section, we describe basic ideas underlying the various tools we will use for the analysis presented in this paper. We begin with a discussion of quantum channels and instruments, continue with local operations and classical communication (LOCC), followed by a discussion of infinite-round LOCC, and finally, the geometric objects we will need, known as zonoids.
	
	\subsection{Quantum Channels, Measurements, and Instruments}\label{sec21}
	Let $\HC$ be a Hilbert space of finite dimension $d$. Denote the set of bounded linear operators acting on $\HC$ as $\BC(\HC)$. A quantum channel $\EC$ is a completely positive, trace-preserving (CPTP) linear map which sends density operators $\rho\in\BC(\HC)$ to density operators $\EC(\rho)\in\BC(\HC_o)$, where a density operator $\rho$ is positive semidefinite and has trace equal to unity. Any channel $\EC$ may be represented as follows. Consider a trio of Hilbert spaces, $\HC,\HC_o,\HC_e$, and an isometry $V$, so that $V^\dag V=I_{\HC}$, $V:\HC\to\HC_o\otimes\HC_e$, with $I_{\HC}$ the identity operator on $\HC$. This maps operator $\rho\in\BC(\HC)$ to operator $V\rho V^\dag\in\BC(\HC_o)\otimes\BC(\HC_e)$. $\HC$ describes the input to the channel, $\HC_o$ the output, and $\HC_e$ is commonly referred to as the environment, which can be viewed as introducing noise into the evolution of the input system as it evolves to the output. Taking partial traces, we obtain two separate channels,
	\begin{align}
		\label{eqn10}\EC(\rho)&=\pTr{e}{V\rho V^\dag}\notag\\
		\FC(\rho)&=\pTr{o}{V\rho V^\dag}.
	\end{align}
	\noindent Channel $\FC$, which maps density operators on $\BC(\HC)$ to those on $\BC(\HC_e)$, is called the complementary channel with reference to $\EC$, and vice-versa. Note that output and environment spaces for $\FC$ are exchanged relative to those for $\EC$, while the inputs are the same. One can always write
	\begin{align}
		\label{eqn11}V=\sum_{j=1}^{d_e}\ket{j}_e\otimes K_j=\sum_{l=1}^{d_o}\ket{l}_o\otimes Q_l,
	\end{align}
	where $\ket{j}_e\in\HC_e$, while $K_j$ maps states on $\HC$ to states on $\HC_o$ and are referred to as Kraus operators \cite{Kraus} for channel $\EC$. Similarly, $\ket{l}_o\in\HC_o$ while $Q_l:\HC\to\HC_e$ and are Kraus operators for the complementary channel, $\FC$. From \myeq{eqn11}, it is easily seen that $_o\bra{l}K_j=~_e\bra{j}Q_l$. This relation, along with the representation of $\EC$ and $\FC$ by $V$, will be used in Appendix~\ref{secA}.

	In the main text of this paper, we will use an alternative representation for a quantum channel, which may be obtained from \myeq{eqn10} and \myeq{eqn11} in terms of Kraus operators as $\EC(\rho)=\sum_jK_j\rho K_j^\dag$ (and $\FC(\rho)=\sum_lQ_l\rho Q_l^\dag$). The choice of such a representation is not unique. Rather, with $\kappa$ the Kraus rank, which gives the minimum number of operators needed for a Kraus representation of $\EC$, if $\{\hat K_m\}_{m=1}^\kappa$ is a minimal representation, then the set $\{K_j\}$ is also a representation of $\EC$ as long as $K_j=\sum_jW_{jm}\hat K_m$ for every $j$, where $W$ is any isometry, $W^\dag W=I_\kappa$ and $I_\kappa$ is the $\kappa\times\kappa$ identity matrix. There are no constraints on individual Kraus operators, but for channels, as a collection they must satisfy the condition $\sum_jK_j^\dag K_j=V^\dag V=I_{\HC}$. Any given set of Kraus operators also represents a specific measurement, each $K_j$ transforming state $\rho\to K_j\rho K_j^\dag$ with probability $p_j=\Tr{K_j^\dag K_j\rho}$. When there is a classical record providing index $j$, we will refer to the set $\{K_j\}$ as a \emph{generalized measurement}. The collection of operators $\{K_j^\dag K_j\}$ constitutes a positive-operator valued measure (POVM), which arises from a measurement after which the final state of the system is discarded, and each $E_j=K_j^\dag K_j$ is called a POVM element. We draw a distinction between a generalized measurement and a POVM in this way.
	
	A \emph{quantum instrument} is a set of completely positive (but generally not trace-preserving) maps, $\EC=\{\JC_r\}$, so that $\JC_r(\rho)=\sum_{j\in S_r}K_{j}\rho K_{j}^\dag$, and $S_r$ is a subset of the full index set identifying those $K_j$ that are associated with CP map $\JC_r$. The collection of these CP maps is trace-preserving, $\sum_r\sum_{j\in S_r}K_{j}^\dag K_{j}=I_\HC$. A generalized measurement is equivalent to a fine-grained quantum instrument having one and only one Kraus operator for each CP map, whereas more general instruments may involve multiple Kraus operators for one or more of its CP maps. For example, from a measurement involving three operators, $K_1,K_2,K_3$, a fine-grained quantum instrument (generalized measurement) is the collection $\JC_r(\rho)=K_r\rho K_r^\dag,~r=1,2,3$. On the other hand, the pair of CP maps, $\JC_1^\prime(\rho)=K_1\rho K_1^\dag+K_2\rho K_2^\dag$ and $\JC_2^\prime(\rho)=K_3\rho K_3^\dag$ together also constitute a quantum instrument, being a coarse-grained version of the former. When the sum over terms is unrestricted, the instrument involves only a single CP map, which is then trace-preserving and is therefore, itself, a quantum channel. On the other hand, every quantum instrument is uniquely associated with a quantum channel through the identification,
	\begin{align}
		\label{eqn12}\EC(\rho)=\sum_r\JC_r(\rho)\otimes[r]_c.
	\end{align}
	The projector $[r]_c$ is to be understood as a register holding the classical information about the various CP maps, since measuring this register in the $\ket{r}_c$ basis provides knowledge of which of these CP maps was implemented. These channels are referred to as \emph{quantum-classical} maps.
	
	\subsection{Local Operations and Classical Communication}\label{sec22}
	 In this paper, we are concerned with quantum instruments on multipartite systems, where a multipartite system on $P$ parties is described by a tensor product Hilbert space, $\HC=\bigotimes_\alpha\HC_{\alpha}$, $\alpha\in\{A,B,\ldots\}$, and we shall refer to $\HC_\alpha$ as the \emph{local} Hilbert space of party $\alpha$. We wish to study the question of whether or not channel (or instrument) $\EC$, acting on multipartite system $\HC$ can be closely approximated when the parties are restricted to using local operations and classical communication. That is, can parties $A,B,\ldots$ closely simulate $\EC$ by making local measurements in their individual laboratories, communicating the outcomes of those measurements to the other parties, continuing in this way for as many rounds of measurement and communication as they wish, and then at the very end, forgetting the outcomes of those local measurements? To analyze these questions, it is common to represent an LOCC protocol by a tree graph. The root of the tree corresponds to the situation before any party has done anything, so we label this node by the identity operator, $I_\HC$. For each local measurement by any given party, each outcome is represented by a node with an edge extending upward in the tree (in the direction of the root) to the node representing the situation just before that measurement was performed. The latter node is known as the \emph{parent} of its collection of \emph{child} nodes. A finite branch stretches from the root node to a leaf node, which has no children, but a protocol may also have infinite branches, which since they do not end, have no leaf nodes. Since these trees represent LOCC protocols, for which only one party measures at a time, the difference between parent and child is \emph{local}. To see this explicitly suppose, as we have found it productive to do \cite{mySEPvsLOCC,myQChImpossbyLOCC,myStrongBounds,myProdPaths,myProdPathsDiscriminate}, that each node in the tree is labeled by the POVM element corresponding to the action of all parties up to that point in the protocol. That is, if the parent node is labeled as the POVM element $\AC\otimes\BC$ and Alice measures next, for example, then the child node will be $\AC^\prime\otimes\BC$; only Alice's local part has changed. Finally, we recall a lemma from \cite{myLOCCbyFirstMeas}, which applies to all finite-round protocols where each node is labeled by a positive semidefinite operator, as just described.
	\begin{lem9}
		\label{lem9}\cite{myLOCCbyFirstMeas}Each node $n$ in a finite-round LOCC tree is equal to the sum of the collection of all leaf nodes that are descendant from that node $n$.
	\end{lem9}
	
	The foregoing discussion may give the appearance that only POVMs are being considered, but one can show that any generalized measurement consisting of operators $K_j^\prime=U_j\sqrt{K_j^\dag K_j}$, with $U_j^\dag U_j=I_\HC$, can be implemented by essentially the same LOCC protocol as that implementing POVM elements $K_j^\dag K_j$, as long as each $U_j$ is a product operator of the form $U_{jA}\otimes U_{jB}\otimes\ldots$. This is because product operators are the only ones LOCC protocols are capable of implementing, so since $K_j$ and $K_j^\prime$ must both be product operators, then $U_j$ must be one, as well. In order to implement the $K_j^\prime$, the parties simply perform the same protocol as that implementing the $K_j$, after which they add one last step at the end, each party $\alpha$ performing $U_{j\alpha}$ after $\sqrt{K_j^\dag K_j}$ has been obtained (here viewing the original protocol as implementing a generalized measurement rather than a POVM). Since the reverse implication is obvious, that implementation of the $K_j$ implies that the $K_j^\dag K_j$ can also be implemented, we see that as long as all the $U_j$ are product unitaries, then the generalized measurement can be implemented by LOCC if and only if the corresponding POVM can be so implemented. A couple of other points are worth noting. First, even though a quantum channel can be represented by many alternative sets of Kraus operators, any given LOCC protocol implements a specific such set, so may equally well be viewed as having implemented the particular measurement corresponding to that set. We will use this observation below and will refer to each corresponding measurement $\MC$ as being \emph{compatible} with the given channel (or instrument) $\EC$. Secondly, we note that the outcome of each local measurement in an LOCC protocol must be communicated to the other parties, and any given implementation of the protocol will follow a single branch with a string of local outcomes leading to a given final outcome, where these outcomes have all been communicated and therefore, recorded. Technically, however, when the outcome is known, one has failed to implement the channel. This is why, in discussing LOCC implementation of quantum channels, we imagine that at the end of the protocol, the parties \emph{forget} the collection of outcomes obtained in each of the individual local measurements. Similarly, for implementing quantum instruments, other than the fully fine-grained measurements, one forgets all classical information other than index $r$ identifying CP map $\JC_r$.
	
	\subsection{Asymptotic LOCC and convergence of quantum channels}\label{sec23}
	To study what happens with \emph{asymptotic LOCC}, where an infinite number of rounds is allowed, one considers the infinite-round protocol to be the limit of a given sequence of finite-round protocols. As discussed in Ref.~\cite{WinterLeung}, there are two distinct kinds of such sequences. In the first case, each protocol in the sequence is obtained from the protocol immediately preceding it by adding one (or more) additional rounds. Such protocols are still LOCC. Let us now discuss the other case, in which the next protocol in the sequence again has additional rounds but is also allowed to differ from its preceding protocol even in earlier rounds. As such, apart from the assumption of convergence, there need not be a specific relationship between successive protocols of the sequence. With either type of asymptotic LOCC, if the instruments implemented by the sequence of protocols have a limiting instrument, then we say that the latter is in, or can be implemented by, the closure of LOCC, denoted $\locc$. That is, if protocol $\PC^\nu$ implements instrument $\EC^\nu$ and $\lim_{\nu\to\infty}\EC^\nu=\EC$, then we shall say that instrument $\EC$ can be implemented by $\locc$: $\EC\in\locc$. What these limits mean is that for every real $\epsilon>0$ there exists natural number $\mu$ such that for every natural number $\nu>\mu$,
	\begin{align}\label{eqn101}
		\normm{\EC-\EC^\nu}_\diamond<\epsilon,
	\end{align}
	\noindent where the diamond norm distance is defined as \begin{align}
		\label{eqn120}\normm{\EC-\EC^\nu}_\diamond=\max_{\normm{X}_1\le1}\normm{(\IC\otimes\EC)X-(\IC\otimes\EC^\nu)X}_1,
	\end{align}
	with $\IC$ the identity channel on $\BC(\HC)$, $X$ is an operator on $\BC(\HC)\otimes\BC(\HC)$, and $\normm{\cdot}_1$ is the trace norm, $\normm{X}_1=\Tr{\sqrt{X^\dag X}}$. A lower bound for this distance may be obtained in terms of these channels' Choi-Jamiolkowski operators, $\Phi$ and $\Phi^\nu$, respectively. This is done by choosing a maximally entangled state on $\BC(\HC)\otimes\BC(\HC)$ in \myeq{eqn120}, $X=\ket{\Omega}\bra{\Omega}=\sum_{ij}\dyad{i}{j}\otimes\dyad{i}{j}/d$, and noting the definition of the Choi-Jamiolkowski operator for $\EC$ as $\Phi=(\IC\otimes\EC)\ket{\Omega}\bra{\Omega}$, with an identical relation between $\Phi^\nu$ and $\EC^\nu$ ($\EC$ and $\EC^\nu$ must act on the same input Hilbert space if a sequence of the latter is to converge to the former). %\frac{1}{d}\sum_{i,j=1}^{d}\sum_{m=1}^{d_e}\ket{i}\bra{j}\otimes K_m\ket{i}\bra{j}K_m^\dag$, with $d$ the dimension of $\HC$.
	Inserting this in \myeq{eqn101} yields,
	\begin{align}\label{eqn102}
		\normm{\Phi-\Phi^\nu}_1<\epsilon.
	\end{align}
	Therefore, convergence of the channels to $\EC$ implies convergence of the Choi-Jamiolkowski operators  to $\Phi$.
				
	\subsection{Zonotopes and Zonoids}\label{sec24}
	We will make use of certain geometric objects known as zonoids, which are limits of objects known as zonotopes. A zonotope is a closed, convex set, the Minkowski sum of a finite number, $N$, of line segments, $[0,E_j]$, written $Z=\sum_j[0,E_j]$. An alternative definition is $Z=\left\{z\left\vert z=\sum_jc_jE_j,0\le c_j\le1\right.\right\}$. Zonoids are limits of sequences of zonotopes as $N\to\infty$. We have utilized zonotopes, $Z_\MC$, generated by POVM elements $E_j\in\MC$, so that the $E_j$ are positive semidefinite operators in a Hilbert space, $\BC(\HC)$, the sum of which is equal to $I_\HC$ \cite{myProdPaths,myProdPathsDiscriminate}. The zonoids, $Z_\EC$, used here will be generated by a minimal Kraus representation, $\{\hat K_m\}$, of a quantum channel $\EC$ in the sense that
	\begin{align}\label{eqn201}
		Z_\EC=\left\{z\left\vert z=\sum_{mm^\prime}\hat C_{mm^\prime}\hat K_m^\dag\hat K_{m^\prime},I_\kappa\ge\hat C\ge0\right.\right\},
	\end{align}
	\noindent with $\hat C$ a positive semidefinite $\kappa\times\kappa$ matrix, $\kappa$ is the Kraus rank of channel $\EC$, and $I_\kappa$ is the $\kappa\times\kappa$ identity matrix. Since positive semidefinite matrices can always be diagonalized, we can write $\hat C_{mm^\prime}=\sum_jW_{jm}^\ast c_jW_{jm^\prime}$ for some isometry, $W$, so that $z$ in the above definition of $Z_\EC$ becomes
	\begin{align}\label{eqn202}
		z&=\sum_jc_j\left(\sum_{m}W_{jm}^\ast\hat K_m^\dag\right)\left(\sum_{m^\prime}W_{jm^\prime}\hat K_{m^\prime}\right)=\sum_jc_jK_j^\dag K_j=\sum_jc_jE_j,
	\end{align}
	Since \myeq{eqn202} holds for every set of coefficients, $\{c_j\}$ and every isometry $W$, we see that if $z\in Z_\MC$, then $z\in Z_\EC$ as well, and we have thus proved the following lemma.	
	\begin{lem6}\label{lem6}
	Given zonoid $Z_{\EC}$ as defined in \myeq{eqn201} and zonotope $Z_{\MC}$ for any $\MC$ compatible with $\EC$ in the sense that measurement operators $K_j$ of $\MC$ constitute a Kraus representation of $\EC$, then $Z_\MC\subseteq Z_\EC$.
	\end{lem6}
	
	The following two examples may provide the reader with a bit of intuition about these geometric objects. First, consider a simple quantum channel on a single qubit, $d=2$, represented by Kraus operators that are rank-$1$ projectors, $\{[0],[1]\}$, with $[\psi]=\ket{\psi}\bra{\psi}$. Operators $K_l=V_{l0}[0]+V_{l1}[1]$, with $l=1,\ldots,N$ for any $N$, also represent this same quantum channel, as long as $V^\dag V=I_2$, the $2\times2$ identity matrix. Then, for any $z\in Z_\EC$, we have that 
	\begin{align}
		\label{eqn111} z=C_{00}[0] + C_{11}[1].
	\end{align}
	Since $[0][1]$ vanishes, there is no contribution from the off-diagonal entries of $C$. Thus, $Z_\EC$ is generated by a finite number (two) of line segments, these segments stretching from the zero operator to $[0]$ and $[1]$, respectively. Therefore, $Z_\EC$ is in fact a zonotope, rather than a zonoid, and has the shape of a square.   
	
	The second example is again on a single qubit, in this case represented by the two Kraus operators $\dyad{0}{0},\dyad{0}{1}$. In this case, any set of operators, $\dyad{0}{\psi_l}$, satisfying $\sum_l[\psi_l]=I_2$ also represents this same channel. Any $z\in Z_\EC$ is of the form
	\begin{align}
		\label{eqn112} z=C_{00}[0]+C_{01}\dyad{0}{1}+C_{10}\dyad{1}{0}+C_{11}[1],
	\end{align}
	showing that similarly to $C$, the only constraint on $z$ is that $0\le z\le I_\HC$. Then, $Z_\EC$ is a four-dimensional cone-like object which, denoting the fourth dimension as $r=\Tr{z}$, stretches from $0\le r\le2$, and every three-dimensional slice through $Z_\EC$ and perpendicular to the $r$-axis is a (Bloch-like) sphere of radius $r$. Given these spherical cross-sections, $Z_\EC$ can only be generated by $N$ line segments in the limit of $N\to\infty$, so this object is a true zonoid.
	
	\subsection{Notation}
	A heads-up for the reader about notation: superscripts that are Greek letters, such as $\mu,\nu$, are not exponents, but are labels denoting position in a sequence; see for example, superscript $\nu$ in the various equations of Section~\ref{sec23}, above. Those superscripts that are lowercase letters from the Roman/Latin alphabet, such as $n$, represent exponents. Subscripts are always just labels, and all lowercase Roman/Latin labels may be assumed to be natural numbers.
	
	\section{Approximating quantum channels by LOCC}\label{sec3}
	In this section we present our main result, a necessary condition that $\EC\in\locc$, Theorem~\ref{thm1}, the proof of which is given in the subsequent section. We begin by recalling the main theorem of Ref.~\cite{myProdPaths}. There, we proved that if measurement $\MC=\{E_j\}$ can be approximated by LOCC arbitrarily closely, then for each $j$, there exists a continuous, monotonic path, $\Pi_j(s)$, of positive semidefinite product operators from $I_\HC$ to each outcome $E_j$ of that measurement and lying entirely within the zonotope, $Z_{\MC}$, generated by those POVM elements, $\{E_j\}$. Here, monotonicity means that the trace of $\Pi_j(s)$ is non-increasing along the path. The proof of that result relied on the observation that every finite-round LOCC protocol implements a measurement for which there exists a continuous, monotonic, \emph{piecewise local}\footnote{The notion of a piecewise local path is borrowed from the piecewise constant curves we all learned about when studying calculus. Instead of being constant, however, the pieces here correspond to \emph{local} measurement outcomes and correspond to the edges between a parent node and one of its children in an LOCC tree. Along this edge, only one party's part of the full measurement operator changes, justifying use of the word local. See Ref.~\cite{myProdPaths} for a more detailed explanation of this concept.} path of positive semidefinite product operators to each of the outcomes of that measurement. Using this observation, it was then argued that from any sequence of finite-round LOCC protocols implementing measurements $\MC^\nu$ for which $\lim_{\nu\to\infty}\MC^\nu=\MC$, there exists a sequence of piecewise local paths whose limit is a (not necessarily piecewise local) path lying within $Z_{\MC}$ and ending at one of the outcomes of $\MC$, and that this holds for each such outcome. The resulting theorem reads as follows.
	\begin{thm2}\cite{myProdPaths}\label{thm2}
		If ${\cal M}\in\overline{\textrm{LOCC}}$, with measurement ${\cal M}$ consisting of POVM elements $E_j$, then for each $j$, there exists a continuous, monotonic path of product operators from ${\cal I}_{\cal H}$ to a point on the (half-open) line segment $(0,E_j]$, and this path lies entirely within zonotope $Z_{\cal M}=\sum_j[0,E_j]$.
	\end{thm2}
	\noindent As argued above, the theorem holds whether one views $\MC$ as a POVM or as a generalized measurement. In either case, it is the POVM elements $E_j$ that determine zonotope $Z_\MC$ and the endpoints of the consequent paths. 
	
	In this section, we are concerned with quantum channels rather than measurements. Since any LOCC protocol implementing channel $\EC$ may just as well be viewed as implementing the generalized measurement consisting of a specific set of Kraus operators which represents $\EC$, Theorem~\ref{thm2} provides a solid foundation for the following theorem, which is our main result, providing a necessary condition that a quantum channel may be simulated arbitrarily closely by LOCC.
	\begin{thm1}\label{thm1}
		Given quantum channel $\EC$ acting on input space $\BC(\HC)$ and represented by the minimal set of $\kappa$ Kraus operators, $\{\hat K_m\}$, then if $\EC$ can be implemented by LOCC with vanishingly small error, the following conditions must hold:
		\begin{enumerate}
			\item\label{itm1} There exists a set, $\{\Pi_j(s)\}$, of continuous, monotonic paths of positive semidefinite product operators, each of which begins at $I_\HC$, the identity operator on $\HC$, and ends at a positive semidefinite product operator of the form, $E_j=\sum_{mm^\prime}(\hat C_j)_{mm^\prime}\hat K_m^\dag\hat K_{m^\prime}$.
			\item\label{itm2} Each $\hat C_j$ is a positive semidefinite matrix, $\hat C_j\ge0$, and may be chosen to have rank equal to unity, with the collection satisfying $\sum_j\hat C_j=I_\kappa$, where $I_\kappa$ is the $\kappa\times\kappa$ identity matrix and $\kappa$ is the Kraus rank of $\EC$. 
			\item\label{itm3} Each of these paths of product operators lies entirely within the zonoid, $Z_{\EC}$, consisting of all positive semidefinite linear combinations of the operators, $\hat K_m^\dag\hat K_{m^\prime}$: $Z_{\EC}=\left\{z\left\vert z=\sum_{mm^\prime}\hat C_{mm^\prime}\hat K_m^\dag\hat K_{m^\prime}, I_\kappa\ge\hat C\ge0\right.\right\}$, with $1\le\textrm{rank}(\hat C)\le\kappa$.% \cmmt{might there be confusion between these $C$'s here and the limiting $C$'s?? No, I think the limiting $C$'s are $C_j$'s . . . }
		\end{enumerate}
	\end{thm1}
	\noindent Note that a \emph{positive semidefinite linear combination} simply means a linear combination of the form given in the above definition of $Z_{\EC}$, that is, a linear combination where the coefficients constitute a positive semidefinite matrix. In addition, the paths are parametrized by $s=\Tr{\Pi_j(s)}$, whose range is $d\ge s\ge0$ with $d$ the dimension of input Hilbert space $\HC$; see \cite{myProdPaths} for a more detailed explanation of this parametrization.
	
	The condition of Theorem~\ref{thm1} that $\EC$ can be implemented by LOCC with vanishingly small error means there exists a sequence of finite-round protocols $\PC^\nu$ implementing quantum channel $\EC^\nu$ such that $\lim_{\nu\to\infty}\EC^\nu=\EC$. Since as discussed in Section~\ref{sec22}, each of these finite-round protocols may equally well be viewed as implementing a measurement $\MC^\nu$, which implies the existence of the concomitant paths of product operators in Theorem~\ref{thm2} \cite{myProdPaths}, this (at least) suggests the existence of the set of limiting paths alluded to in Theorem~\ref{thm1}. Thus, one should not be surprised to find that the results for approximating measurements by LOCC \cite{myProdPaths} can be generalized, as we do here, to apply in similar fashion to approximating quantum channels by LOCC. One notable difference was discussed in Section~\ref{sec24}, that while the approximation of measurements involves paths lying entirely within a zonotope, $Z_{\MC}$, for the approximation of channels one must utilize the zonoid, $Z_{\EC}$.
	
	Recall from Section~\ref{sec24} that in order to implement a quantum channel, say $\EC^\nu$, by LOCC, one may implement any measurement $\MC^\nu$ compatible with $\EC^\nu$. This means that simply knowing one can implement a given channel leaves one without knowledge of which compatible measurement(s) can be implemented. Thus, we also do not know the set of Kraus operators implemented, whereas that set is what determines the endpoints of the paths introduced in Theorem~\ref{thm2} \cite{myProdPaths}. Therefore, we are left with the question, how do we know what the endpoints of all the paths of Theorem~\ref{thm1} are? The answer is, we do not. Instead, as is indicated in Item~\ref{itm2} of Theorem~\ref{thm1}, we only know that each of these endpoints corresponds to a rank-$1$ matrix $\hat C_j$ and the collection of these matrices constitutes a rank-$1$ decomposition of the $\kappa\times\kappa$ identity matrix. This is the extent to which we are able to characterize those endpoints without knowing in advance what specific set of Kraus operators is to be implemented. Nonetheless, this is sufficient to ensure that the resulting set of Kraus operators are a representation of $\EC$.
	
	We now turn to the proof of Theorem~\ref{thm1}.
	
\section{Proof of Theorem~\ref{thm1}}\label{sec4}
	We begin the proof of Theorem~\ref{thm1} with the following lemma showing that convergence of a sequence of quantum channels implies convergence of the corresponding sequence of zonoids.
	\begin{lem2}\label{lem2}
		If there exists a sequence of quantum channels, $\EC^\nu$ of Kraus rank $\kappa^\nu$ and with minimal Kraus representation $\{\hat K_m^\nu\}$, converging to channel $\EC$ of Kraus rank $\kappa$ and with minimal Kraus representation $\{\hat K_m\}$---$\lim_{\nu\to\infty}\EC^\nu=\EC$---then there also exists a sequence of zonoids, $Z_{\EC^\nu}=\left\{z^\nu\left\vert z^\nu=\sum_{mm^\prime}\hat C^\nu_{mm^\prime}\hat K_m^{\nu\dag}\hat K_{m^\prime}^\nu,I_{\kappa^\nu}\ge\hat C^\nu\ge0\right.\right\}$ associated with channel $\EC^\nu$, which converges to zonoid $Z_\EC=\left\{z\left\vert z=\sum_{mm^\prime}\hat C_{mm^\prime}\hat K_m^{\dag}\hat K_{m^\prime},I_{\kappa}\ge\hat C\ge0\right.\right\}$ associated with channel $\EC$. That is, $\lim_{\nu\to\infty}Z_{\EC^\nu}=Z_\EC$.
	\end{lem2}
	\noindent The proof is given in Appendix~\ref{secA}.
	
	The next step is to recall that each LOCC protocol $\PC^\nu$, which implements channel $\EC^\nu$, may also be thought of as implementing a POVM, $\MC^\nu$, consisting of elements $E_j^\nu=K_j^{\nu\dag}K_j^\nu$, where with $W^\nu$ an isometry, $K_j^\nu=\sum_mW_{j m}^\nu\hat K_m^\nu$ is a Kraus operator in a collection representing channel $\EC^\nu$. Therefore, by Theorem~\ref{thm2}, we know that there exists a continuous, monotonic path of product operators from $I_\HC$ to a point on the half-open line segment $(0,E_j^\nu]$, for each outcome, $j$, and that these paths lie entirely within the zonotope $Z_{\MC^\nu}$ generated by the collection of these operators, ${E_j^\nu}$. In fact, by Theorem $2$ of Ref.~\cite{myProdPaths}, these paths are piecewise local. As discussed at the end of the preceding section, we do not know $\MC^\nu$.	Therefore, we  consider all possible $\MC^\nu$ compatible with channel $\EC^\nu$, which means that instead of considering our continuous paths as lying within zonotope $Z_{\MC^\nu}$, we instead view these paths as lying within zonoid $Z_{\EC^\nu}$. According to Lemma~\ref{lem6}, $Z_{\MC^\nu}\subseteq Z_{\EC^\nu}$ for every such $\MC^\nu$, so this is always possible. We could, of course, restrict to measurements that are implementable by LOCC, so that such paths of product operators actually exist, but it is sufficient to follow the easier route and include \emph{all} compatible measurements, as we do here.
	
	Thus, we have a sequence of zonoids, $\{Z_{\EC^\nu}\}$, which converges to zonoid $Z_\EC$, and within each $Z_{\EC^\nu}$, we have a collection of continuous paths of product operators starting at $I_\HC$. This situation is very similar to the one encountered in Ref.~\cite{myProdPaths}, including that the paths are all Lipschitz continuous with Lipschitz constant equal to unity \cite{myProdPaths}. By the Arz\'ela-Ascoli theorem \cite{ArzelaAscoli}, this tells us that there exists a subsequence of these paths which converges uniformly on the range, $s\in[d,0]$, to a continuous limiting path. However, to this point, we have many paths for each $\nu$ and have not actually identified a sequence of paths to begin with. We now turn to the task of identifying such a sequence. In fact we will identify, and characterize, many such sequences.
	
	Whereas in our earlier work, we had a sequence of known measurements, $\MC^\nu$, whose limit was another known measurement, $\MC$, here we do not know what the $\MC^\nu$ or $\MC$ are. Instead, we only know that each $\MC^\nu$ is compatible with channel $\EC^\nu$, the sequence of which has the limit $\EC$. In the previous case, knowing $\MC^\nu$ meant that we also knew the final outcomes, $E_j^\nu$, of the corresponding LOCC protocol, and also that there was an ordering of these outcomes, provided by labels $\nu$ for fixed $j$, such that for each $j$, the sequence $\{E_j^\nu\}_\nu$ satisfied $\lim_{\nu\to\infty}E_j^\nu=E_j$, with $E_j$ an outcome of the limiting measurement $\MC$. This allowed us to identify sequences of paths---i.e., those ending at $E_j^\nu$ with the aforementioned fixed label, $j$---from which, according to Arz\'ela-Ascoli, it followed that there was a limiting path from $I_\HC$ to a point along $(0,E_j]$. In our present case, with $\lim_{\nu\to\infty}\EC^\nu=\EC$, we do not know $E_j^\nu$ or the limiting outcomes $E_j$. Thus, we need a way to partition the various paths we've identified as lying within the $Z_{\EC^\nu}$ so that each subset provides a sequence of paths whose limit we can characterize. In particular, we wish to characterize the endpoints of the limiting paths. We will discuss how to partition the paths forthwith. The following lemma concerning the Kraus ranks of the $\EC^\nu$ will be useful, and it may be of interest in other contexts, as well.
	\begin{lem5}\label{lem5}
		Consider the sequence of quantum channels, $\{\EC^\nu\}$, with the Kraus rank of $\EC^\nu$ being $\kappa^\nu$. If $\lim_{\nu\to\infty}\EC^\nu=\EC$ with Kraus rank of $\EC$ equal to $\kappa$, then there exists $\mu$ such that $\kappa^\nu\ge\kappa$ for all $\nu>\mu$.
	\end{lem5}
	\proof We begin the proof by considering the Choi-Jamiolkowski operator $\Phi$ $(\Phi^\nu)$ for $\EC$ $(\EC^\nu)$, defined below \myeq{eqn120}. It is well known that the rank of $\Phi$ is equal to the Kraus rank $\kappa$ of $\EC$, and similarly for $\Phi^\nu$. By the Eckart-Young-Mirsky low rank approximation theorem \cite{EckartYoungMirsky}, which holds for all unitarily invariant norms \cite{MirskyUInvariant}, we have that when rank$(\Phi^\nu)<\kappa$, then $\normm{\Phi-\Phi^\nu}_1\ge\sigma_{\kappa}$, where $\sigma_{\kappa}$ is the $\kappa$th largest singular value of $\Phi$. Note that since $\EC$, and thus $\Phi$, is known, $\sigma_\kappa$ is some fixed, positive number. Choose $\epsilon$ to satisfy $\sigma_{\kappa}>\epsilon>0$ and then identify $\mu$ such that for every $\nu>\mu$, $\epsilon>\normm{\EC^\nu-\EC}_\diamond$. Since $\lim_{\nu\to\infty}\EC^\nu=\EC$, this is always possible.
	
	The proof of the lemma now proceeds by contradiction.  Suppose for some $\nu>\mu$,  $\kappa^\nu<\kappa$. Recalling from the discussion around Eqs.~(\ref{eqn101}) and (\ref{eqn102}) that $\normm{\EC-\EC^\nu}_\diamond\ge\normm{\Phi-\Phi^\nu}_1$, then for every $\epsilon<\sigma_{\kappa}$, we have
	\begin{align}\label{eqn31}
		\sigma_{\kappa}>\epsilon>\normm{\EC-\EC^\nu}_\diamond\ge\normm{\Phi-\Phi^\nu}_1\ge\sigma_{\kappa},
	\end{align}
	a contradiction. Therefore, either $\nu<\mu$ or $\kappa^\nu\ge\kappa$. That is to say, for each $\epsilon>0$, there exists $\mu$ such that $\kappa^\nu\ge\kappa$ for all $\nu>\mu$, and the proof is complete.\endproof
	
	\noindent The meaning of this result is clear: a sequence of low rank operators cannot become arbitrarily close to an operator of higher rank. On the other hand, it is clearly possible for higher rank operators to become closer and closer to an operator of lower rank.
	
	We now turn to a characterization of endpoints of our paths. To begin, notice that each branch in finite-round LOCC protocol $\PC^\nu$ ends at a leaf node corresponding to POVM element $E_j^{\nu}$, or equivalently at measurement operator, $K_j^{\nu\prime}=V^\nu\sqrt{E_j^{\nu}}=\sum_lW_{j l}^\nu\hat K_l^\nu$, where $V^\nu$ and $W^\nu$ are isometries (acting on different spaces, of course) and the $\hat K_l^\nu$ constitute a minimal Kraus representation of the quantum channel $\EC^\nu$ implemented by $\PC^\nu$. Then,
	\begin{align}\label{eqn30}
		E_j^{\nu}&=\sum_{ll^\prime}W_{j l}^{\nu\ast}W_{j l^\prime}^\nu \hat K_l^{\nu\dag}\hat K_{l^\prime}^\nu\notag\\
		&=\sum_{ll^\prime}(\hat C_j^\nu)_{ll^\prime}\hat K_l^{\nu\dag}\hat K_{l^\prime}^\nu,
	\end{align}
	with $\hat C_j^\nu=\vec W_j^\nu\vec W_j^{\nu\dag}$ a rank-$1$ matrix, and $\vec W_j^\nu$ is the $j$th column of isometry $W^\nu$. Note that from Ref.~ \cite{myProdPaths}, we know that each branch corresponds directly to a continuous monotonic path of product operators lying entirely within $Z_{\MC^\nu}$, and thus within $Z_{\EC^\nu}$. We see here that each of these paths ends at a point corresponding to a rank-$1$ matrix, $\hat C_j^\nu$.
	
	Let us consider the collection of these endpoints and the paths $\Pi_j^\nu(s)$ to each of them. Following Ref.~\cite{myProdPaths}, we parametrize these paths by $s=\Tr{\Pi_j^\nu(s)}$. Define $N^\nu$ to be the number of these paths, equal to the number of leaf nodes, $j=1,\ldots,N^\nu,$ for each protocol $\PC^\nu$. By the results of Ref.~\cite{myStrongBounds}, we can assume that $N^\nu$ is finite for all $\nu<\infty$. Consider the first $\lambda$ of these protocols, $\nu=1,\ldots,\lambda$, and set $N_0(\lambda)=\max_{\nu\le\lambda}{N^\nu}$. For each $\nu$ such that $N^\nu<N_0(\lambda)$, split up some of the paths into $p_j^\nu$ duplicates. We can do this splitting at the very final (local) measurement of each path that is split. For example, if the final outcome of $\Pi_j^\nu(s)$ is $E_j^\nu$, replace that single outcome by $p_j^\nu$ individual outcomes, each of which is $E_j^\nu/p_j^\nu$. This increases the number of outcomes of that final local measurement by $p_j^\nu-1$. It doesn't matter which final outcomes are duplicated, or how many parts we choose to split each of them into, as long as we end up with a new number of paths such that $N^{\nu\prime}=N_0(\lambda)$ for each $\nu$. Note that each of these paths correspond to a branch in an LOCC protocol, $\PC^{\nu\prime}$, which is only slightly modified from a branch in $\PC^\nu$ by splitting a subset of the final outcomes as just described. Redefine each of these final segments of the duplicated paths as $\Pi_j^{\nu\prime}(s)=\Pi_j^\nu(s)/p_j^\nu$, for $s$ in the appropriate range for that final local measurement (the paths are not altered for the remainder of the range of $s$). This maintains the condition that the sum of  POVM elements corresponding to the entire collection of leaf nodes for protocol $\PC^{\nu\prime}$ is equal to $I_\HC$, as it must be. We now have that each protocol $\PC^{\nu\prime}$ has $N_0(\lambda)$ final outcomes, with a continuous path $\Pi_j^{\nu\prime}(s)$ to each one, and each of these final outcomes corresponds to a positive semidefinite operator satisfying \myeq{eqn30} with matrix $\hat C_j^\nu$ also positive semidefinite and having rank equal to unity. Since the sum of all final outcomes is equal to $I_\HC$, there is always a choice of these matrices $\hat C_j^\nu$ such that they are a rank-$1$ decomposition of the identity, $I_{\kappa^\nu}$ (if, for example, the $K_l^{\nu\dag}K_{l^\prime}^\nu$ are linearly dependent, then there can be other possible choices for the $\hat C$-matrices, but we are specifically looking for those that are rank-$1$).
	
	For each $j=1,\ldots,N_0(\lambda)$, we have $\lambda$ paths, $\Pi_j^\nu(s),~\nu=1,\ldots,\lambda$ (dropping the primes, for convenience).	Since $N_0(\lambda)$ is a maximum over the first $\lambda$ protocols, then $N_0(\lambda^\prime)\ge N_0(\lambda)$ whenever $\lambda^\prime>\lambda$; $N_0(\lambda)$ is a non-decreasing function of $\lambda$. When incrementing $\lambda\to\lambda+1$, one path is added for each value of $j\le N_0(\lambda)$. If $N_0(\lambda+1)>N_0(\lambda)$, additional values of $j$ are then added (by splitting outcomes as discussed in the preceding paragraph) up to $N_0(\lambda+1)$, along with $\lambda+1$ paths for each new value of $j$, one from each of the protocols $\PC^\nu$ for $\nu=1,\ldots,\lambda+1$. It is possible that $\hat N_0=\lim_{\lambda\to\infty}N_0(\lambda)$ diverges in the limit $\lambda\to\infty$ (see Section~\ref{sec5} for a detailed analysis of such a case) but in any event, we end up with an infinite number of paths for each $j=1,\ldots,\hat N_0$ in this limit. For each path, we have a rank-$1$ matrix $\hat C_j^\nu$ associated with its endpoint. We will now argue that for each $j$, the sequence of sets of matrices $\{\hat C_j^\nu\}_\nu$ has a subsequence with a limit $\{\hat C_j\}$ the collection of which is also a rank-$1$ decomposition of the identity. This argument is somewhat complicated by the fact that $\kappa^\nu$ can vary as $\nu$ increases.
	
	For each $j$, we apply the Arzel\`a-Ascoli theorem \cite{ArzelaAscoli} (see Ref.~\cite{myProdPaths} for details applicable to the present context) to obtain a limiting path, but we make these limits for different $j$ consistent with each other as follows. For $j=1$ there exists a subsequence, $S_1=\{\nu_{11},\nu_{12},\ldots\}$, of paths $\Pi_1^{\nu_{1j}}(s)$ for which, by Arzel\'a-Ascoli, a limiting path of product operators exists, denote it $\Pi_1(s)$. Since the endpoint of each path in this subsequence of paths is associated with a rank-$1$ matrix $\hat C_1^{\nu_{1j}}$, then by the Eckart-Young-Mirsky theorem \cite{EckartYoungMirsky}, the endpoint of the limiting path must also be represented by a rank-$1$ matrix $\hat C_1$. Next, for $j=2$, consider only those paths $\Pi_2^{\nu}(s)$ and matrices $\hat C_2^{\nu}$ such that $\nu\in S_1$. Here, again by Arzel\'a-Ascoli, there exists a subsequence $S_2=\{\nu_{21},\nu_{22},\ldots\}$ of $S_1$, for which the corresponding set of paths has a limiting path of product operators, $\Pi_2(s)$, and by the same arguments used above for $S_1$, the endpoint of this limiting path is again represented by a rank-$1$ matrix $\hat C_2$. Since $S_2$ is a subsequence of $S_1$, we have that $\lim_{\nu\to\infty,\nu\in S_2}\Pi_1^\nu(s)=\Pi_1(s)$, and $\lim_{\nu\to\infty,\nu\in S_2}\hat C_1^\nu=\hat C_1$; these limits remain the same. Continuing this process for each $j$, we find a subsequence $S_j$ of $S_{j-1}$ along with a limiting path $\Pi_j(s)$ and rank-$1$ matrix $\hat C_j$. Furthermore, the limiting paths and matrices obtained with subsequence $S_{j^\prime}$ continue to be valid limits when restricted to $S_j$, for all $j^\prime<j$. Since for each $\nu$, $\sum_j \hat C_j^\nu=I_{\kappa^\nu}$, one might na\"ively expect that the set of limiting matrices $\hat C_j$ satisfies a similar condition, $\sum_j \hat C_j=I_{\kappa}$. This conclusion is (essentially) correct, but let us make the argument with care.

	As discussed above, rank-$1$ matrices $\hat C_j^\nu$ obey the condition, $\lim_{\nu\to\infty}\hat C_j^\nu=\hat C_j$, and their sum behaves as $\lim_{\nu\to\infty}\sum_j \hat C_j^\nu=\lim_{\nu\to\infty}I_{\kappa^\nu}$. And what is the limit of a sequence of identity operators? Note that $I_{\kappa}-I_{\kappa^\nu}$ can never be small unless $\kappa^\nu=\kappa$. From the proof of Lemma~\ref{lem5}, we know that $\kappa^\nu>\kappa$ is possible, so this equality may not hold. Nonetheless, we must recognize that this is not the relevant limit for discussing convergence of a sequence of quantum channels implemented by LOCC protocols. Rather, as we have shown above, what is relevant are the operators, $E_j^\nu=\sum_{mm^\prime}(\hat C_j^\nu)_{mm^\prime}\hat K_m^{\nu\dag}\hat K_{m^\prime}^\nu$. Now, $\sum_j E_j^\nu=\sum_{mm^\prime}(I_{\kappa^\nu})_{mm^\prime}\hat K_m^{\nu\dag}\hat K_{m^\prime}^\nu=\sum_m\hat K_m^{\nu\dag}\hat K_m^\nu=I_\HC$, as required. It may happen that as $\nu$ becomes large, the norm of $\hat K_m^\nu$ may become negligible, tending asymptotically to zero. In that case, even though $\sum_j \hat C_j^\nu=I_{\kappa^\nu}$, the only non-negligible parts of the sum for $E_j^\nu$ (see a few sentences previous to this one) correspond to matrix elements $(\hat C_j^\nu)_{mm^\prime}$ for which $\normm{\hat K_m^\nu},\normm{\hat K_{m^\prime}^\nu}$ do \emph{not} tend to zero. Under these circumstances, while $\sum_j \hat C_j^\nu=I_{\kappa^\nu}$, it is only the part of $I_{\kappa^\nu}$ lying in that subspace corresponding to the non-negligible $\hat K_m^\nu$ that is relevant. Since by assumption the limit of this sequence of channels exists and is equal to $\EC$ of Kraus rank $\kappa$, then according to these arguments that relevant subspace must be of size $\kappa$. By Lemma~\ref{lem5}, we have $\kappa^\nu\ge\kappa$ for large enough $\nu$, which guarantees the existence of a subspace of size $\kappa$ within the  $\kappa^\nu$-dimensional space upon which $I_{\kappa^\nu}$ acts, for large enough $\nu$. Therefore, $I_{\kappa^\nu}$ may (essentially) be replaced by the projector onto this subspace, which is (essentially) equivalent to $I_{\kappa}$. With this, the proof of Theorem~\ref{thm2} is complete.\endproof
	
	\section{Detailed example of a quantum channel}\label{sec5}
	In this section, we consider the example \cite{WinterLeung} of a two-qubit state transformation which can be arbitrarily closely approximated, but cannot be implemented exactly by LOCC. In Section~\ref{sec7}, we consider a generalization of this example to any number of qubits. Consider a \emph{quantum instrument} \cite{WinterLeung}, which consists of the following three CP maps, the sum of which form the quantum channel $\EC$ we consider here, but are not individually trace preserving,
	\begin{align}\label{eqn300}
		\JC_1(\rho)&=[11]\rho[11]\notag\\
		\JC_2(\rho)&=\sum_{i=1}^{2}(T_i\otimes[0])\rho(T_i^\dag\otimes[0])\notag\\
		\JC_3(\rho)&=\sum_{i=1}^{2}([0]\otimes T_i)\rho([0]\otimes T_i^\dag),
	\end{align}
	where $T_1=I/\sqrt{3}$ and $T_2=\left([0]+2[1]\right)/\sqrt{6}$. This suggests describing channel $\EC$ by the five product Kraus operators,
	\begin{align}\label{eqn301}
		K_{1}^\prime&=[11]\notag\\
		K_{2}^\prime&=\sqrt{\frac{1}{3}}\left([00]+[10]\right)\notag\\
		K_{3}^\prime&=\sqrt{\frac{1}{6}}\left([00]+2[10]\right)\notag\\
		K_{4}^\prime&=\sqrt{\frac{1}{3}}\left([00]+[01]\right)\notag\\
		K_{5}^\prime&=\sqrt{\frac{1}{6}}\left([00]+2[01]\right).
	\end{align}
	It turns out that there is a smaller set of (product) Kraus operators that represent the same channel. For the sake of simplicity, we will use this smaller set, which is
	\begin{align}\label{eqn302}
		\hat K_1&=[11]\notag\\
		\hat K_2&=\frac{2}{3}[00]+[10]\notag\\
		\hat K_3&=\frac{2}{3}[00]+[01]\notag\\
		\hat K_4&=\frac{1}{3}[00].
	\end{align}
	It is easily checked that $K_j^\prime=\sum_jW_{jl}\hat K_l$, with $W$ the isometry,
	\begin{align}\label{eqn303}
		W=
		\begin{pmatrix}
			1&0&0&0\\
			0&\sqrt{\frac{1}{3}}&0&\sqrt{\frac{1}{3}}\\
			0&\sqrt{\frac{2}{3}}&0&-\sqrt{\frac{1}{6}}\\
			0&0&\sqrt{\frac{1}{3}}&\sqrt{\frac{1}{3}}\\
			0&0&\sqrt{\frac{2}{3}}&-\sqrt{\frac{1}{6}}
		\end{pmatrix}.
	\end{align}

\begin{figure}
		\centering
		\includegraphics[width=\textwidth]{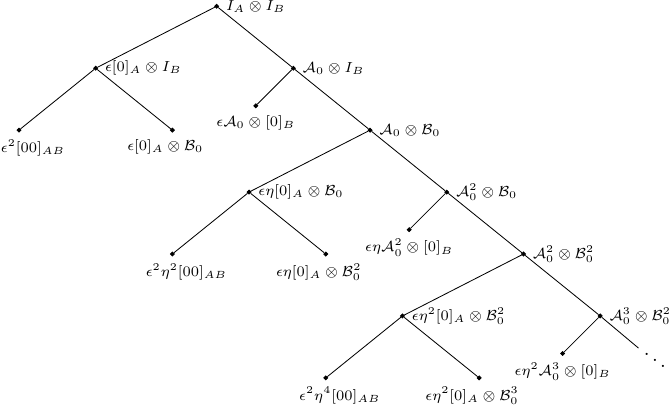}
		\caption{\label{fig1}Tree graph for the first several local measurements performed by Alice and Bob in any one of the sequence of LOCC protocols described in the text. Note that we have defined $\eta=1-\epsilon$.}
	\end{figure}

	The following LOCC protocol \cite{FortescueLo,WinterLeung}, $\PC^\nu$, implements a quantum channel, $\EC^\nu$, which approximates the channel $\EC$ of \myeq{eqn302}. Given single qubit (Hermitian) operators, $A_0=\sqrt{1-\epsilon}[0]_A+[1]_A$ and $A_1=\sqrt{\epsilon}[0]_A$, Alice begins by measuring $\{A_0,A_1\}$ on her system $A$, with corresponding POVM elements $\AC_j=A_j^\dag A_j=A_j^2,~j=0,1$. She shares her outcome with Bob, who follows with the same measurement---having POVM elements $\BC_0=(1-\epsilon)[0]_B+[1]_B$ and $\BC_1=\epsilon[1]_B$---on his system $B$, sharing his outcome in turn with Alice. They halt at this point unless they both obtained outcome $0$, in which case they repeat their local measurements as just described. Then, after a maximum of $\nu$ iterations of this procedure, they stop. Noting that $A_0^n=(1-\epsilon)^{n/2}[0]_A+[1]_A$ and $A_1A_0^n=\epsilon^{1/2}(1-\epsilon)^{n/2}[0]_A$, and similarly for Bob's measurement operators, then the first few rounds of this LOCC protocol---represented as a tree graph with nodes labeled by the corresponding POVM elements implemented to that point in the protocol---is depicted in Figure~\ref{fig1} (note that the notation is condensed slightly in this figure by defining $\eta=1-\epsilon$). In the figure, there is a single main branch, stretching down along the upper right side, which continues (note ellipses at bottom right) for $2\nu$ segments, while all other branches extend either one or two segments off of that main branch and then terminate. As discussed in \cite{WinterLeung}, the limiting channel is $\EC$, as desired, as long as we choose $\epsilon=\nu^{-c}$ with $0<c<1$, which we will do in what follows.
	
	Consider that main branch in each protocol and the sequence of these main branches formed as we increment $\nu$. Each of these main branches describes a path, $\Pi^{\nu}(s)$, of product operators in operator space, as described in Sections~\ref{sec3} and \ref{sec4}. The sequence of such paths of product operators has a limiting path, which we will now identify. Defining $s=\Tr{\AC_0^n\otimes\BC_0^n}=[(1-\epsilon)^n+1]^2$, we have $\sqrt{s}-1=(1-\epsilon)^n$, at least at those particular nodes along this main branch. This suggests the quantity $\sqrt{s}-1$ may play a role. Noting that we will be considering the limit as $\epsilon\to0$, and that in this limit, we will also have the length of the branch, $\nu\to\infty$, care must be taken in considering how the quantity $(1-\epsilon)^n$ behaves in these limits (since $n$ ranges up to $\nu$). Nonetheless, the quantity $\sqrt{s}-1$ is perfectly well-behaved in these limits, and in fact, we can prove the following Lemma.
	\begin{lem1}\label{lem1}
		The limiting path of product operators obtained from the sequence of paths, $\Pi^{\nu}(s)$, is $\Pi_M(s)=\lim_{\nu\to\infty}\Pi^{\nu}(s)=M(s)\otimes M(s)$, $4\ge s\ge1$, with $M(s)=\left(\sqrt{s}-1\right)[0]+[1]$.
	\end{lem1}
	\noindent Note that $\Tr{\Pi_M(s)}=s$, so the path is parametrized by the trace, as has been done in \cite{myProdPaths}. 
	\proof 
	Define $t_n=(1-\epsilon)^n+1$, $s_{2n}=\Tr{\AC_0^n\otimes\BC_0^n}=t_n^2$ and $s_{2n+1}=\Tr{\AC_0^{n+1}\otimes\BC_0^n}=t_nt_{n+1}$. For each $\nu,\epsilon$, and $0\le n\le\nu$, consider a piecewise local segment from $\AC_0^{n}\otimes\BC_0^n$ to $\AC_0^{n+1}\otimes\BC_0^n$ along the main branch of the $\nu$th protocol, with $s_{2n}\ge s\ge s_{2n+1}$ or $t_n^2\ge s\ge t_nt_{n+1}$. This segment along the path may be parametrized as
	\begin{align}\label{eqn305}
		\Pi^{\nu}(s)&=\frac{s-s_{2n+1}}{s_{2n}-s_{2n+1}}\AC_0^n\otimes\BC_0^n+\frac{s_{2n}-s}{s_{2n}-s_{2n+1}}\AC_0^{n+1}\otimes\BC_0^n\notag\\
				&=\frac{1}{t_n(t_n-t_{n+1})}\left[(s-t_nt_{n+1})
				\begin{pmatrix}
					t_n-1&0\\
					0&1
				\end{pmatrix}
				+(t_n^2-s)
				\begin{pmatrix}
					t_{n+1}-1&0\\
					0&1
				\end{pmatrix}
				\right]\otimes
				\begin{pmatrix}
					t_n-1&0\\
					0&1
				\end{pmatrix}\notag\\
				&=\frac{1}{t_n}
				\begin{pmatrix}
					s-t_n&0\\
					0&t_n
				\end{pmatrix}
				\otimes
				\begin{pmatrix}
					t_n-1&0\\
					0&1
				\end{pmatrix}.
	\end{align}
	We need to show that for each $s$ and for all $\delta>0$, there exists $\mu$ such that for all $\nu>\mu$, $\delta>\normm{\Pi^{\nu}(s)-\Pi_M(s)}_1$. Consider
	\begin{align}\label{eqn306}
		t_n\bra{00}\left[\Pi^{\nu}(s)-\Pi_M(s)\right]\ket{00}&=(s-t_n)(t_n-1)-t_n(\sqrt{s}-1)^2\notag\\
							&=2t_n\sqrt{s}-t_n^2-s\notag\\
							&\le2t_n\sqrt{s_{2n}}-t_n^2-s_{2n+1}=t_n^2-t_nt_{n+1}=t_n(t_n-t_{n+1}).
	\end{align}
	Since $t_n-t_{n+1}=\epsilon(1-\epsilon)^n$, this is small. Similarly,
	\begin{align}\label{eqn307}
		t_n\bra{01}\left[\Pi_M(s)-\Pi^{\nu}(s)\right]\ket{01}&=t_n(\sqrt{s}-1)-(s-t_n)\notag\\
					&=t_n\sqrt{s}-s\notag\\
					&\le t_n(t_n-t_{n+1}),
	\end{align}
	and it is easy to see that $\bra{10}\left[\Pi^{\nu}(s)-\Pi_M(s)\right]\ket{10}=t_n-\sqrt{s}\le t_n-t_{n+1}$, as well. Since the $11$ diagonal element vanishes identically, we have that
	\begin{align}\label{eqn308}
		\normm{\Pi^{\nu}(s)-\Pi_M(s)}_1\le\sqrt{3}(t_n-t_{n+1})=\sqrt{3}\epsilon(1-\epsilon)^n\le\sqrt{3}\epsilon=\sqrt{3}\nu^{-c}.
	\end{align}
	Considering the segments from $\AC_0^{n+1}\otimes\BC_0^n$ to $\AC_0^{n+1}\otimes\BC_0^{n+1}$, corresponding to $s_{2n+1}\ge s\ge s_{2n+2}$, one may follow the same argument as above---the only difference being that $t_{n+1}\to t_{n+2}$ and $t_n\to t_{n+1}$---yielding the exact same result. Since this encompasses all cases, then for the entire path along that main branch, we may choose $\mu=\left(\delta/\sqrt{3}\right)^{-1/c}$ yielding $\normm{\Pi^{\nu}(s)-\Pi_M(s)}_1<\delta$ for all $\nu>\mu$ and for all $s$, and this ends the proof.\endproof
	
	It is instructive to understand what happens with the other outcomes of our sequence of protocols. Recalling that $s=[(1-\epsilon)^n+1]^2$ and that $\epsilon$ decreases as $\nu$ increases, the value of $s$ at which the $n$th side-branch emerges from the main branch increases with $\nu$. One may envision this as these side-branches sliding up the main branch toward the root node in Fig.~\ref{fig1} as $\nu$ increases, as more and more of these side-branches are added further down the main branch. In the limit $\nu\to\infty$, the gap between each side-branch and its nearest neighbors vanishes, and we have a continuous set of complete separable (generally, not LOCC\footnote{Note that since the limiting main branch is not piecewise local---that is, each point along this branch differs from the next in both party's local operators, not one party's at a time---these measurements at each $s$ are not LOCC.}) measurements being made, one at each $s$, with one outcome being $\Pi_M(s-\Delta s)$ and the other outcomes we will now determine. By taking the limit as $\Delta s\to0$, we see that it is simply neccessary to differentiate the path along the main branch. This gives
	\begin{align}\label{eqn309}
		d\Pi_M(s)=\frac{ds}{2\sqrt{s}}\left([0]\otimes M(s)+M(s)\otimes[0]\right),
	\end{align}
	which is not a product operator, as it must be for the limiting channel to be in $\locc$. However, we can simply split it into two parts, each of which is clearly a product operator, providing two additional outcomes for each of the continuous set of measurements appearing at the various values of $s$. Notice the path from $\Pi_M(s)$ to either one of these parts is, in fact, piecewise local, extending from $M(s)\otimes M(s)$ to either $d\sqrt{s}[0]\otimes M(s)$ or $d\sqrt{s}M(s)\otimes[0]$. One of these outcomes clearly corresponds to those of the form $\epsilon(1-\epsilon)^n[0]\otimes\BC_0^n$ in each of our sequence of LOCC protocols, and the other to $\epsilon(1-\epsilon)^n\AC_0^n\otimes[0]$. One may conjecture that the finite-round parts of (convergent) sequences of LOCC protocols always converge to piecewise local paths, as we have now seen happens for this specific example, which also demonstrates how the parts whose number of rounds tends to infinity can converge to paths that are not piecewise local. Note that in the limit $\epsilon\to0$, there are no paths to outcomes proportional to $[00]$, consistent with the observation (see Fig.~\ref{fig1}) that in each of the sequence of protocols, these outcomes are proportional to $\epsilon^2$ and so become negligible. On the other hand, outcomes proportional to $\epsilon$ are not negligible, since their number diverges as $\nu=\epsilon^{-1/c}$, and taken together, they provide a significant contribution. To summarize then, the POVM implemented at the end of the protocol illustrated in Fig.~\ref{fig1} in the limit as $\nu\to\infty$ ($\epsilon\to0$) consists of the following operators,
	\begin{align}
		\label{eqn340}E_1&=[11]\notag\\
						E_2(s)&=d\sqrt{s}M(s)\otimes[0],~4\ge s\ge1\notag\\
						E_3(s)&=d\sqrt{s}[0]\otimes M(s),~4\ge s\ge1.
	\end{align}
	
	From the foregoing discussion, we also see that in the limit $\Delta s\to0$, we end up with a continuum of measurements, one for each value of $s$ ranging from $s=4$ down to $s=1$. That is,
	\begin{align}
		\label{eqn3401}M(s)\otimes M(s)\to\{M(s-ds)\otimes M(s-ds),\frac{ds}{2\sqrt{s}}[0]\otimes M(s), \frac{ds}{2\sqrt{s}}M(s)\otimes[0]\},
	\end{align}
	which may be written in a perhaps more enlightening way as
	\begin{align}
		\label{eqn3402}M(s)\otimes M(s)\to\{M(s-ds)\otimes M(s-ds),ds\frac{dM(s)}{ds}\otimes M(s), dsM(s)\otimes\frac{dM(s)}{ds}\}.
	\end{align}
	Notice that outcome $M(s-ds)\otimes M(s-ds)$ cannot be obtained from $M(s)\otimes M(s)$ via a local measurement. Thus, this is not an LOCC protocol. Nonetheless, we do have a sequence of measurements, so we see that this limit does lead to an (infinite) protocol. It is just not an LOCC protocol, but rather, one that is non-local.
	
	From \myeq{eqn340}, one immediately obtains the limiting generalized measurement operators, for example $K_2(s)=\sqrt{E_2(s)}$, when viewing the sequence of protocols as implementing generalized measurements rather than a POVM. That is,
	\begin{align}
		\label{eqn341}K_1&=[11]\notag\\
		K_2(s)&=\sqrt{d\sqrt{s}M(s)\otimes[0]},~4\ge s\ge1\notag\\
		K_3(s)&=\sqrt{d\sqrt{s}[0]\otimes M(s)},~4\ge s\ge1.
	\end{align}
	
	We now calculate the associated Choi-Jamiolkowski operator to demonstrate that this infinite set of limiting Kraus operators represents the desired channel $\EC$, see \myeq{eqn302}. There is one limiting Kraus operator obtained at the end of the main branch in the limit, which is $\sqrt{M(1)\otimes M(1)}=[11]=\hat K_1$. In addition, within the range $4\ge s\ge1$, we have two infinite sets of Kraus operators, $K_2(s)$ and $K_3(s)$.
	
	Define $\sigma=\sqrt{s}-1$ so that $d\sqrt{s}=d\sigma$, and let the Choi-Jamiolkowski operator be written as $\Omega=\Omega_1+\Omega_2+\Omega_3$, with
	\begin{align}\label{eqn330}
		\Omega_1=\sum_{w,x,y,z=0}^{1}\ket{wx}\bra{yz}^{A^\prime B^\prime}\otimes \hat K_1\ket{wx}\bra{yz}^{AB}\hat K_1=[1111],
	\end{align}
	\begin{align}\label{eqn331}
		\Omega_2&=\sum_{w,x,y,z=0}^1\ket{wx}\bra{yz}^{A^\prime B^\prime}\otimes\int_{1}^{2}d\sqrt{s}\left(\sqrt{M(s)}\otimes[0]\right)\ket{wx}\bra{yz}^{AB}\left(\sqrt{M(s)}\otimes[0]\right)\notag\\
			&=\int_1^2d\sigma\left(\sqrt{\sigma}\ket{0000}+\ket{1010}\right)\left(\sqrt{\sigma}\bra{0000}+\bra{1010}\right)\notag\\
			&=\frac{1}{2}[0000]+\frac{2}{3}\left(\ket{0000}\bra{1010}+\ket{1010}\bra{0000}\right)+[1010],
	\end{align}
	and in entirely similar fashion, we find that
	\begin{align}\label{eqn332}
		\Omega_3&=\frac{1}{2}[0000]+\frac{2}{3}\left(\ket{0000}\bra{0101}+\ket{0101}\bra{0000}\right)+[0101].
	\end{align}
	Adding these together yields
	\begin{align}\label{eqn333}
		\Omega&=I_{\HC\otimes\HC}+\frac{2}{3}\left(\ket{0000}\bra{0101}+\ket{0101}\bra{0000}+\ket{0000}\bra{1010}+\ket{1010}\bra{0000}\right)`,
	\end{align}
	in agreement with the results of \cite{WinterLeung} for the Choi-Jamiolkowski operator of $\EC$.\footnote{There seem to be typographical errors in \cite{WinterLeung} in their calculation of this Choi-Jamiolkowski operator---we believe their $A^\prime A$ matrices should be $4\times4$ instead of $2\times2$---but their intended meaning is clear.}
	
	Another way to demonstrate that these limiting Kraus operators represent the desired channel $\EC$ is by showing that they are related to the set in \myeq{eqn302} by an isometry. In this case, the isometry will have an infinite number of rows. We have that one of these Kraus operators is $\hat K_1$, so that the first row of our isometry is $[1~0~0~0]$. The other rows are obtained from $K_2(s)$ and $K_3(s)$, written as
	\begin{align}\label{eqn707}
		K_2(s)&=\sqrt{d\sigma M(s)\otimes[0]}=\sqrt{d\sigma}\left(\hat K_2+(3\sqrt{\sigma}-2)\hat K_4\right),\notag\\
    	K_3(s)&=\sqrt{d\sigma[0]\otimes M(s)}=\sqrt{d\sigma}\left(\hat K_3+(3\sqrt{\sigma}-2)\hat K_4\right).
	\end{align}
	That is, we obtain a continuous set of additional rows from each of these: $\sqrt{d\sigma}[0~1~0~(3\sqrt{\sigma}-2)]$ and $\sqrt{d\sigma}[0~0~1~(3\sqrt{\sigma}-2)]$. As an isometry, all columns must be orthonormal. The first column is $[1~0~0~\ldots~0]^T$, so this column is normalized as well as orthogonal to the other columns, since each of their first entries is zero. Normalization of the second and third columns follows from the fact that $\int_{1}^{2}d\sigma=1$. Orthogonality of this pair of columns is clear, since the non-zero entries in one are always zero in the other. Normalization of the last column is
	\begin{align}\label{eqn708}
		2\int_{1}^{2}d\sigma(3\sqrt{\sigma}-2)^2&=2\int_{0}^{1}d\sigma(9\sigma -12\sqrt{\sigma}+4)=1,
	\end{align}
	where the factor of $2$ is included because there are two entries in this column for each value of $s$, one from $K_2(s)$ and the other from $K_3(s)$, see \myeq{eqn707}. Thus, we see that all columns are normalized. Orthogonality of the last column to either of the middle two is
	\begin{align}\label{eqn709}
		\int_{0}^{1}d\sigma(3\sqrt{\sigma}-2)=0
	\end{align}
	Thus, we do indeed see that this continuous set of Kraus operators is related to those in \myeq{eqn302} by an isometry.
	
	As noted in Item~\ref{itm2} of Theorem~\ref{thm1}, the endpoints of our limiting paths should be described by positive semidefinite matrices which have rank equal to unity. The current example illustrates this aspect, as well. We have that $d\sigma M(s)\otimes[0]=\sum_{mn} C_2(s)_{mn}\hat K_m^\dag \hat K_n$, with the $\hat K_m$ given in \myeq{eqn302} and
	\begin{align}\label{eqn310}
		 C_2(s)=d\sigma
		\begin{pmatrix}
			0&0&0&0\\
			0&1&0&3\sqrt{\sigma}-2\\
			0&0&0&0\\
			0&3\sqrt{\sigma}-2&0&\left(3\sqrt{\sigma}-2\right)^2
		\end{pmatrix}.
	\end{align}
	Similarly, $d\sigma[0]\otimes M(s)=\sum_{mn} C_3(s)_{mn}\hat K_m^\dag \hat K_n$, with
	\begin{align}\label{eqn311}
		 C_3(s)=d\sigma
		\begin{pmatrix}
			0&0&0&0\\
			0&0&0&0\\
			0&0&1&3\sqrt{\sigma}-2\\
			0&0&3\sqrt{\sigma}-2&\left(3\sqrt{\sigma}-2\right)^2
		\end{pmatrix}.
	\end{align}	
	Both $C_2(s)$ and $C_3(s)$ are rank-$1$ matrices for every $s$. They also provide an easy way to find a matrix $C_1(s)$ from which $\Pi_M(s)=\sum_{mn} C_1(s)_{mn}\hat K_m^\dag \hat K_n$, the path along the (limit of the) main branch in Fig.~\ref{fig1},  may be obtained. That is, since the collection of these matrices must sum to $I_\kappa$, we have
	\begin{align}\label{eqn312}
		 C_1(s)&=I_4-\int_{\sqrt{s}}^{2}\left( C_2(s)+ C_3(s)\right)\notag\\
		&=\begin{pmatrix}
			1&0&0&0\\
			0&0&0&0\\
			0&0&0&0\\
			0&0&0&0
		\end{pmatrix}
		+\sigma
		\begin{pmatrix}
			0&0&0&0\\
			0&1&0&2\left(\sqrt{\sigma}-1\right)\\
			0&0&1&2\left(\sqrt{\sigma}-1\right)\\
			0&2\left(\sqrt{\sigma}-1\right)&2\left(\sqrt{\sigma}-1\right)&9\sqrt{s}-1-16\sqrt{\sigma}
		\end{pmatrix}.
	\end{align}
	This is equal to the identity matrix at $\sigma=1$, of course, and since the second term in the second line of \myeq{eqn312} vanishes at $\sigma=0$, it leads to the endpoint of the limiting main branch being described by $\Pi_M(1)=\hat K_1^\dag \hat K_1$, with $\textrm{rank}\left(C_1(1)\right)=1$, as required. Furthermore, one calculates using the set of operators in \myeq{eqn302} that $\sum_{mn}(C_1(s))_{mn}\hat K_m^\dag\hat K_n=M(s)\otimes M(s)$ for all $s$, as it must.

	Note that $C_2(s)$ and $C_3(s)$ correspond to the endpoints of side branches that begin at position $s$ of the main branch. It is easily seen that the entire length of these side branches may be represented by matrices $C_2(s,x)=(1-x)C_1(s)+xC_2(s)$ and $C_3(s,x)=(1-x)C_1(s)+xC_3(x)$, $0\le x\le1$. Since these are convex combinations of positive semidefinite matrices, they are themselves positive semidefinite for all $s$ and $x$.
	
	According to the definition of $Z_\EC$ in Item~\ref{itm3} of Theorem~\ref{thm2}, the fact that $C_1(s)$ and $C_j(s,x),~j=2,3,$ are all positive semidefinite matrices shows that the paths associated with all these limiting measurement outcomes lie entirely within $Z_\EC$. It is not difficult to see that most of the length of the paths associated with the main branch of the individual LOCC protocols in our approximating sequence do not lie in $Z_\EC$, but only approach it in the limit. While the endpoints of those approximating side branches do lie within $Z_\EC$, they begin at the main branch so are at least partly outside of $Z_\EC$, as well.
	
	\section{Quantum instruments}\label{sec9}
	Generalized measurements and quantum channels share similar characteristics, both involving transformations of state $\rho$ to $K_j\rho K_j^\dag$, with the channel involving an unrestricted sum over the latter quantities, there being no accessible classical information about the outcome $j$. Both are special cases of the more general notion of a a quantum instrument, which covers circumstances where some classical information is available but possibly less than is required for a full generalized measurement. The proof of Theorem~\ref{thm1} for channels in Section~\ref{sec4} is also capable of encompassing all quantum instruments as long as we make allowances for this required classical information, which can be embedded in the form of the positive semidefinite $C$-matrices introduced in Theorem~\ref{thm1}, as we now discuss.
	
	In the definition we have given here for zonoid $Z_\EC$, sums over terms of the form $\hat C_{mm^\prime}\hat K_m^\dag\hat K_{m^\prime}$ appear. On the other hand, in our analysis of quantum measurements \cite{myProdPaths,myProdPathsDiscriminate}, the zonotopes used there involved simple, positive linear combinations of $K_m^\dag K_m$. In other words, for measurements, the $C$-matrices are diagonal, $\hat C_{mm^\prime}=c_m\delta_{mm^\prime}$. We will now argue that a simple generalization to block diagonal matrices will suffice for all quantum instruments.
	
	Recall that to account for all possible Kraus representations of $\EC$, we started with a positive linear combination, $z=\sum_jc_jK_j^\dag K_j$ with $\{K_j\}$ constituting an arbitrary such representation, and then utilized isometric freedom, $K_j=\sum_mW_{jm}\hat K_m$, to obtain $z=\sum_{mm^\prime}\hat C_{mm^\prime}\hat K_m^\dag\hat K_{m^\prime}$. Suppose we have a quantum instrument $\EC$ consisting of CP maps, $\JC_r$, for some set of indices, $r$. To understand what happens to isometric freedom in this case, consider as illustration instrument $\EC=\{\JC_1,\JC_2\}$, with $\JC_1(\rho)=K_{1}\rho K_{1}^\dag+K_{2}\rho K_{2}^\dag$ and $\JC_2(\rho)=K_{3}\rho K_{3}^\dag+K_{4}\rho K_{4}^\dag$. If an LOCC protocol has a final outcome $K=\mu_1K_{1}+\mu_3K_{3}$, this will mix the two CP maps, in general. This is so unless $K_{1}=\alpha K_{3}+\beta K_{4}$ (or $K_{3}=\alpha^\prime K_{1}+\beta^\prime K_{2}$), in which case $K$ can be written as a linear combination of only those Kraus operators involved in expressing $\JC_2$ (or $\JC_1$), and then the parties can obtain enough classical information to identify $K$ as contributing to one of the CP maps and not the other. If such a separation between the CP maps is impossible for one or more outcomes of the protocol, then there is no way to coarse-grain the collection of outcomes such that each subset exactly represents one of those CP maps. Thus for a successful implementation, it must be possible to express each of the final outcomes in a way that does not mix the different CP maps of the desired instrument $\EC$. As the above example illustrates, any linear dependence of the full set of Kraus operators allows for multiple ways of expressing any given outcome, but nonetheless, there must be a choice of these expressions which keeps the different CP maps of $\EC$ separate.
	
	Thus, we see that the full isometric freedom of quantum channels is no longer in effect in this case, but it is still allowed within each $\JC_r$. That is, for example, if $\EC=\{\JC_r\}$ and $\JC_r(\rho)=\sum_{m\in S_r}\hat K_{m}\rho\hat K_{m}$, with $\{\hat K_{m}\}_{m\in S_r}$ a minimal set of Kraus operators for $\JC_r$, we must be able to write any outcome $K_j$ in terms of one, and only one, of these sets. That is for some fixed $r$, $K_j=\sum_{m\in S_r}(W_r)_{jm}\hat K_{m}$ with $W_r^\dag W_r=I_{\kappa_r}$ the $\kappa_r\times \kappa_r$ identity matrix and $\kappa_r$ is the Kraus rank of $\JC_r$. Note that this leads to $K_j^\dag K_j=\sum_{m,m^\prime\in S_r}\hat C_{mm^\prime}\hat K_{m}^\dag\hat K_{m^\prime}$ with $\hat C_{mm^\prime}=(W_r)_{jm}^\ast(W_r)_{jm^\prime}$, so that $C$ is a rank-$1$ matrix, as required by Theorem~\ref{thm1}. In addition, since these outcomes are to be coarse-grained (collected together) to represent $\JC_r$, it must be that
	\begin{align}
		\label{eqn801}\sum_{j\in S_r^\prime} K_j\rho K_j=\sum_{m\in S_r}\hat K_{m}\rho\hat K_{m}=\JC_r(\rho)
	\end{align}
	for some fixed $r$ and index set $S_r^\prime$. This equality follows directly from our identification of $W_r$ as an isometry acting on a $\kappa_r$-dimensional space.
	
	Defining the direct sum, $W=\oplus_r W_r$, we have that $W$ is a block-diagonal isometry mapping the collection of \emph{minimal} (within each $\JC_r$) Kraus sets to a complete set of Kraus operators $\{K_j\}$ for the full protocol implementing instrument $\EC$. Since by Lemma~\ref{lem9} we have that each node in a finite-round LOCC protocol is a positive linear combination of the $K_j^\dag K_j$, then each of these nodes---and therefore, every point along our piecewise local paths identified from these protocols---can be written as a positive semidefinite combination of the $\hat K_{m}^\dag\hat K_{m^\prime}$ with the corresponding $C$-matrix a block diagonal matrix, being itself a sum of block diagonal (rank-$1$) matrices. Thus, the proof of Theorem~\ref{thm1} holds for the case of quantum instruments, except that there is this additional constraint that the $C$-matrices be block diagonal. These matrices keep the different CP maps separate from each other, precluding any mixing between the Kraus operators for $\JC_r$ and those for $\JC_{r^\prime}$ when $r\ne r^\prime$.
	
	These intuitive ideas, which are formalized in Appendix~\ref{secB}, lead directly to a generalization of Theorem~\ref{thm1}, encompassing the most general notion of quantum instruments. Note that to generalize the proof of Lemma~\ref{lem2} for this case, see Appendix~\ref{secB} for details, we simply add this block diagonal constraint on isometries $W,W^\nu$ introduced just before---and matrices $C,C^\nu$ appearing in---\myeq{eqnA17}, and the rest of the proof of this lemma goes through as before.
	\begin{thm8}\label{thm8}
		Given quantum instrument $\EC=\{\JC_r\}$ acting on input space $\BC(\HC)$, with each $\JC_r$ represented by the minimal set of $\kappa_r$ Kraus operators, $\{\hat K_{m}\}_{m\in S_r}$, then if $\EC$ can be implemented by LOCC with vanishingly small error, the following conditions must hold:
		\begin{enumerate}
			\item\label{itm21} There exists a set, $\{\Pi_j(s)\}$, of continuous, monotonic paths of positive semidefinite product operators, each of which begins at $I_\HC$, the identity operator on $\HC$, and ends at a positive semidefinite operator of the form, $E_j=\sum_{mm^\prime}(\hat C_j)_{mm^\prime}\hat K_m^\dag\hat K_{m^\prime}$.
			\item\label{itm22} Each $\hat C_j$ is a positive semidefinite matrix, $\hat C_j\ge0$, and may be chosen to have rank equal to unity, with the collection satisfying $\sum_j \hat C_j=I_\kappa$, where $I_\kappa$ is the $\kappa\times\kappa$ identity matrix and $\kappa=\sum_r\kappa_r$.
			\item\label{itm23} Each of these paths of product operators lies entirely within the zonoid, $Z_{\EC}$, consisting of all positive semidefinite linear combinations of the operators, $\hat K_m^\dag\hat K_{m^\prime}$: $Z_{\EC}=\left\{z\left\vert z=\sum_{mm^\prime}\hat C_{mm^\prime}\hat K_m^\dag\hat K_{m^\prime}, I_\kappa\ge C\ge0\right.\right\}$, with $1\le\textrm{rank}(C)\le\kappa$ and $C$ is block diagonal, the $r$th diagonal block having size $\kappa_r\times\kappa_r$.
			\item\label{itm24} There exists a partition of the index set $j$ into subsets $S_r^\prime$ such that $\sum_{j\in S_r^\prime}\hat C_j=P_{\kappa_r}$, with $P_{\kappa_r}$ a $\kappa\times\kappa$ matrix having the $\kappa_r\times\kappa_r$ identity matrix embedded within its $r$th block and everything else being zero. That is, every matrix element of $P_{\kappa_r}$ vanishes except that the diagonal elements within the $r$th block are equal to unity.
		\end{enumerate}
	\end{thm8}
	
	To illustrate this theorem, we return to the example of the preceding section. The Kraus operators $K_1,K_2(s),K_3(s)$ of \myeq{eqn341} can be written in terms of those of \myeq{eqn301} as
	\begin{align}\label{eqn802}
		K_1&=K_1^\prime\notag\\
		K_2(s)&=\sqrt{d\sigma M(s)\otimes[0]}=\sqrt{3d\sigma}\left[\left(2\sqrt{\sigma}-1\right)K_{2}^\prime+\sqrt{2}\left(1-\sqrt{\sigma}\right)K_{3}^\prime\right],\notag\\
		K_3(s)&=\sqrt{d\sigma[0]\otimes M(s)}=\sqrt{3d\sigma}\left[\left(2\sqrt{\sigma}-1\right)K_{4}^\prime+\sqrt{2}\left(1-\sqrt{\sigma}\right)K_{5}^\prime\right].
	\end{align}
	We see that these Kraus operators, obtained from the limit of our sequence of protocols, are indeed related to those of the quantum instrument, see \myeq{eqn301}, by a block diagonal isometry. The first diagonal block has size $1\times1$, whereas the next two blocks are each infinite in length and two columns wide, and the five columns form an orthonormal set. This implies that the $C$-matrices describing our paths are also block diagonal, as required by Theorem~\ref{thm8}. From another point of view, we may coarse-grain all outcomes $K_2(s)$ together on the one hand, and all outcomes $K_3(s)$ on the other to find
	\begin{align}
		\label{eqn710}\int K_2(s)\rho K_2(s)&=\int_{1}^{2}d\sigma\sqrt{M(s)}\otimes[0]\rho\sqrt{M(s)}\otimes[0]=K_{2}^\prime\rho K_{2}^\prime +K_{3}^\prime\rho K_{3}^\prime=\JC_2(\rho)\notag\\
		\int K_2(s)\rho K_3(s)&=\int_{1}^{2}d\sigma[0]\otimes\sqrt{M(s)}\rho[0]\otimes\sqrt{M(s)}=K_{4}^\prime\rho K_{4}^\prime +K_{5}^\prime\rho K_{5}^\prime=\JC_3(\rho),
	\end{align}
	where we have used \myeq{eqn802}. Thus, recalling that $\JC_1(\rho)=K_1^\prime\rho K_1^\prime$, then by retaining the classical information about which outcome is obtained and when, one is able to coarse-grain the outcomes to separately obtain each of the three CP maps of the full quantum instrument.

	\section{LOCC$\ne\locc$}\label{sec6}
	In this section, we consider the difference between LOCC and $\locc$ for quantum channels and quantum measurements, the extremes of the range of quantum instruments. For quantum channels, we are unable to shed much light on this question. Certainly as we have seen in Section~\ref{sec5}, while the side-branches of the limit of the LOCC protocols implementing channels $\EC^\nu$ are all piecewise local, the limiting main branch, $\Pi_M(s)$, is not. At first blush, one may wonder if this demonstrates that channel $\EC$ cannot be exactly implemented by LOCC, which requires a piecewise local path to each outcome, even if one or more paths may contain an infinite number of pieces. However, it is possible that there is a different set of paths, possibly to the same Kraus representation of $\EC$ or possibly to a different one, all of which are piecewise local. We suspect that this is not the case but have not found a proof.
	
	While we are presently unable to answer this question for quantum channels, we are able to show, we believe for the first time, that LOCC is not closed when considering implementation of quantum measurements, whether the measurements are viewed as POVMs or as generalized measurements. That is, we have the following theorem.
	\begin{thm6}
		\label{thm6}The set of LOCC measurements is not closed.
	\end{thm6}
	\proof We continue with the example of \cite{ChitCuiLoPRL,WinterLeung}, for which we have obtained the limiting POVM, see \myeq{eqn340}, of the sequence of protocols illustrated in Fig.~\ref{fig1}, and the associated limiting generalized measurement, see \myeq{eqn341}. Our measurement operators for the limiting generalized measurement are $[11]$, $\sqrt{d\sigma}[0]\otimes\sqrt{M(s)}$, and $\sqrt{d\sigma}\sqrt{M(s)}\otimes[0]$, and recall that $0\le\sigma\le1$ and $\sigma=\sqrt{s}-1$. We wish to show that this measurement cannot be implemented by LOCC, even though it is the limit of a sequence of LOCC measurements. To show this, we use an argument very similar to the one used in \cite{WinterLeung} for the same sequence of LOCC protocols, but there coarse-grained to the quantum instrument of \myeq{eqn300}. That is, we will consider the action of this generalized measurement---each element of which is tensored with the identity operator on a third party, $C$---on the $W$ state, $\ket{W}=\left(\ket{001}+\ket{010}+\ket{100}\right)/\sqrt{3}$. Then, it is not difficult to see that
	\begin{align}\label{eqn901}
		(K_1\otimes I_C)\ket{W}&=\left([11]\otimes I_C\right)\ket{W}=0\notag\\ (K_2(s)\otimes I_C)\ket{W}&=\sqrt{d\sigma}\left([0]_A\otimes\sqrt{M(s)_B}\otimes I_C\right)\ket{W}=\sqrt{\frac{d\sigma}{3}}[0]_A\otimes\left(\sqrt{\sigma}\ket{01}+\ket{10}\right)_{BC}\notag\\
		(K_3(s)\otimes I_C)\ket{W}&=\sqrt{d\sigma}\left([0]_B\otimes\sqrt{M(s)_A}\otimes I_C\right)\ket{W}=\sqrt{\frac{d\sigma}{3}}[0]_B\otimes\left(\sqrt{\sigma}\ket{01}+\ket{10}\right)_{AC}.
	\end{align}
	The latter pair are a continuum of elements, and what we're interested in are outcomes in the form $\left(K_2(s)\otimes I_C\right)[W]\left(K_2(s)\otimes I_C\right)^\dag$, which we may sum up by integration over $\sigma$. This yields
	\begin{align}\label{eqn902}
		\int_{0}^{1}\left(K_2(s)\otimes I_C\right)[W]\left(K_2(s)\otimes I_C\right)^\dag=[0]_A\otimes\left([01]/3+4\ket{01}\bra{10}/9+4\ket{10}\bra{01}/9+2[10]/3\right)_{BC}
	\end{align}
	with probability equal to one-half, and the exact same thing with the same probability but with parties $A$ and $B$ interchanged for the term involving $K_3(s)$. Given that \myeq{eqn902} is equal to $\left(\JC_2\otimes\IC_C\right)([W])$, compare \myeq{eqn710} (and similarly for $\left(\JC_3\otimes\IC_C\right)([W])$; $\IC_C$ is the identity channel on the Hilbert space, $\HC_C$, representing system $C$), this is in agreement with the result obtained in Section $4.3$ of \cite{WinterLeung}, from which they showed by a concurrence argument \cite{WoottersConcurrence} that this could not be achieved by LOCC. That argument applies here, as well, proving that this is an example of a generalized measurement that is in $\locc$ but not in LOCC. To put it in simple terms, if this measurement could be implemented by LOCC, then by coarse-graining, so could the instrument of \myeq{eqn300}, but this is known to be impossible \cite{WinterLeung}, so the result follows directly.
	
	Note that if the corresponding POVM could be implemented within LOCC, then so could the generalized measurement. Since we have just argued that the latter is impossible, we have also now provided an example of a POVM that is in $\locc$ but not in LOCC, and this ends the proof.\endproof
	
	It is perhaps of interest that this POVM involves an infinite number of outcomes, as does the corresponding generalized measurement. Therefore, one may ask whether or not LOCC is closed for measurements that are restricted to have a finite number of outcomes. If the latter question can be answered in the affirmative, it would show an interesting relationship between the closure properties of LOCC and the number of outcomes allowed in a measurement. The same relationship would then apparently also hold when considering the size of Kraus representation needed to implement a quantum channel or quantum instrument.
	
	Note that these arguments require that one knows which outcomes are in the set, $K_2(s)$, and which are in $K_3(s)$, treating the two sets separately while coarse-graining over the range of $s$. Thus, this example also requires classical information about the outcomes, so it does not apply to the full quantum channel, $\EC$. Therefore, the question of whether or not LOCC is closed for implementing quantum channels remains open.
	
	\section{Generalization of $\EC$ to any number of qubits}\label{sec7}
	Consider a $P$-qubit system with the parties performing the same local measurements as described in Section~\ref{sec5}. That is, each of their local POVM elements are $R=(1-\epsilon)[0]+[1]$ and $I-R$. The parties measure one after the other, and if after any cycle of their $P$ local measurements, any one of them has gotten the second outcome, they stop. Only if they all got $R$ do they continue to the next cycle, and we shall refer to the continued sequence of these cycles as \emph{the main branch} of the protocol and denote it as $\Pi^\nu(s)$, as has been done in Section~\ref{sec5} for the two-qubit case. The parties always stop after $\nu$ rounds, with $\epsilon=\nu^{-c}$ for some $0<c<1$, even if they've always gotten the first outcome, having thus implemented channel $\EC^\nu$. The limiting paths have a form that are a simple generalization of those for the two-qubit case.
	\begin{lem3}
		\label{lem3} In the limit $\nu\to\infty$ (equivalently, $\epsilon\to0$), the main branch of this protocol has a limit given by $\Pi_M(s)=\lim_{\nu\to\infty}\Pi^\nu(s)=M(s)^{\otimes P}$, with $M(s)=(\sqrt[\leftroot{-2}\uproot{2}P]{s}-1)[0]+[1]$.
	\end{lem3}	
	The proof, which requires one to show that $\lim_{\nu\to\infty}\normm{\Pi_M(s)-\Pi^\nu(s)}_1=0$, is straightforward and is left as an exercise for the reader. We share a few hints. Note that after $n$ complete cycles followed by $l$ of the parties having completed their part of the $(n+1)$st cycle, the segment of the (piecewise local) path from this point to the point after the $(l+1)$st party performs the next measurement may be written as
	\begin{align}
		\label{eqn601}\Pi^\nu(s)=\frac{s-t_{n+1}^{l+1}t_n^{P-l-1}}{t_{n+1}^lt_n^{P-l-1}(t_n-t_{n+1})}\left(R^{n+1}\right)^{\otimes l}\left(R^{n}\right)^{\otimes(P-l)}+\frac{t_{n+1}^{l}t_n^{P-l}-s}{t_{n+1}^lt_n^{P-l-1}(t_n-t_{n+1})}\left(R^{n+1}\right)^{\otimes l+1}\left(R^{n}\right)^{\otimes(P-l-1)},
	\end{align}
	with $t_{n+1}^lt_n^{P-l}\ge s\ge t_{n+1}^{l+1}t_n^{P-l-1}$, and $t_n=\Tr{R^n}=\left(1-\epsilon\right)^n+1$. Since all operators are diagonal, the proof of this lemma simply requires one to show that $\norm{\bra{N}\left(\Pi_M(s)-\Pi^\nu(s)\right)\ket{N}}\le\epsilon P$ for every $0\le N\le2^P-1$, where using the binary expansion, $N=b_{P-1}\ldots b_1b_0$ with $b_j=0$ or $1$, we define
	\begin{align}
		\label{eqn604}\ket{N}=\ket{b_0}\ket{b_1}\ldots\ket{b_{P-1}}.
	\end{align}
	Recall that all non-zero local matrix elements are equal to unity unless that party's part of the binary expansion of $N$ is equal to $0$, in which case one obtains a factor of a power of $\left(1-\epsilon\right)$; for example, $\bra{0}R^n\ket{0}=(1-\epsilon)^n=t_n-1$.
	
	As done in the case of $P=2$ above, we may differentiate $\Pi_M(s)$ to obtain other possible Kraus operators for our limiting channel, $\EC_P$. This gives
	
	\begin{align}
		\label{eqn602}d\Pi_M(s)=d\!\!\sqrt[\leftroot{-2}\uproot{2}P]{s}\left([0]\otimes M(s)^{\otimes(P-1)}+M(s)\otimes[0]\otimes M(s)^{\otimes(P-2)}+\ldots+M(s)^{\otimes(P-1)}\otimes[0]\right),
	\end{align}
	suggesting a separable Kraus representation as consisting of $[11\ldots1]$ and each of the individual terms appearing in the preceding equation. In fact, we can show this is correct by calculating the Choi-Jamiolkowski operator for this conjectured channel, as well as the limit of those for the approximating sequence of channels described above in terms of local operators $R$. In the following, $\ket{i},\ket{j}$ are to be understood as a binary expansion, in similar fashion to that for $\ket{N}$ in \myeq{eqn604}, $l_i$ is the number of $0$'s appearing in that binary expansion of $i$, and $m_{ij}$ counts the number of positions where both $i$ and $j$ each have a zero. In other words, $l_i=h(\neg i)$, the Hamming weight of $\neg{i}$ with $\neg$ denoting the bitwise NOT function, and $m_{ij}=h\left(\neg(i|j)\right)$ with $|$ the bitwise OR function. Then, we have
	\begin{lem4}
		\label{lem4}The Choi-Jamiolkowski operator for channel $\EC_P$ and that for the $\nu\to\infty$ limit of the above-described approximating channels, $\EC^\nu$, are both given by
		\begin{align}
			\label{eqn603}\Phi_P=2^{-P}[2^P-1]\otimes[2^P-1]+2^{-P}\sum_{i,j=0}^{2^P-2}\dyad{i}{j}\otimes\frac{2m_{ij}}{l_i+l_j}\dyad{i}{j},
		\end{align}
		and since this is equal to $2^{-P}\sum_{i,j=0}^{2^P-1}\dyad{i}{j}\otimes\EC_P(\dyad{i}{j})$, the corresponding channels are given by
		\begin{align}
			\label{eqn605}\EC_P(\rho)=\bra{2^P-1}\rho\ket{2^P-1}[2^P-1]+\sum_{i,j=0}^{2^P-2}\frac{2m_{ij}\rho_{ij}}{l_i+l_j}\dyad{i}{j},
		\end{align}
		with $\rho_{ij}=\bra{i}\rho\ket{j}$.
	\end{lem4}
	\noindent The proof is given in Appendix~\ref{secC}.
	
	One may consider the action of $\EC_P$ on the state $[W]\otimes[11\ldots1]$ where the first tensor factor is a projector onto the $W$-state of three qubits and the second one is the tensor product of $P-3$ local (qubit) projectors onto the $\ket{1}$ state. Since the action of $M(s)$ on $\ket{1}$ is identical to the action of the identity operator, as is the action of $[1]$, whereas $[0]$ annihilates $\ket{1}$, the last $P-3$ parties' actions are irrelevant to the evolution of the state of the other three parties. It has been shown that the transformation $\EC_3([W])$ cannot be achieved exactly by LOCC \cite{ChitCuiLoPRL}. Therefore, it is also the case that $\EC_P([W])$ cannot be implemented exactly by LOCC---since if it could, then so could $\EC_3([W])$---even though, as we've demonstrated above, LOCC can be used to approximate $\EC_P$ arbitrarily closely. Given the results of \cite{WinterLeung,ChitCuiLoPRL}, these last points are perhaps unsurprising, but the limiting measurement operators in \myeq{eqn602} may be of some interest.

	\section{Conclusions}\label{sec8}
	In this paper, we have extended our earlier result providing a necessary condition that a quantum measurement can be closely approximated by LOCC \cite{myProdPaths,myProdPathsDiscriminate}, to a similar condition which now applies to the full range of quantum instruments from the most refined up to the fully coarse-grained case of quantum channels. The proof of this new result includes a demonstration that if a sequence of quantum instruments converges to some limiting instrument, then there also exists a sequence of geometric objects known as zonoids, which converges to a limiting zonoid. Each such zonoid is uniquely determined by the instrument from which it is derived, and the limiting zonoid plays a critical role in determining whether or not the limiting instrument can be closely approximated by LOCC. When this approximation is possible, there must exist certain paths of product operators lying within the limiting zonoid, see Theorems~\ref{thm1} and \ref{thm8}.
	
	We have studied in detail a particular quantum instrument on two-qubits, $\EC$, which has been shown \cite{WinterLeung} to be the limit of a sequence of instruments, each of which can be implemented by a finite-round LOCC protocol, but that $\EC$ itself cannot be exactly achieved using LOCC. Our analysis shows why this is so. For each of the approximating instruments in the given sequence, there exists a set of piecewise local paths to each of a set of outcomes, where that set of outcomes can be coarse-grained to represent the given instrument. When one considers the limit of these paths, however, one finds that while they correspond to a set of outcomes representing $\EC$, one of the limiting paths fails to remain piecewise local in the limit. Since for LOCC implementation, a crucial characteristic of these paths is that they must be piecewise local, see the discussion above Theorem~\ref{thm2}, this failure is consistent with the finding that $\EC$ cannot be exactly implemented by LOCC \cite{WinterLeung}.
	
	We initially treated this example as a quantum channel, see Section~\ref{sec5}, and found the Kraus representation of $\EC$ corresponding to the limit of the Kraus representations for the approximating  channels $\EC^\nu$. This allowed us to identify the isometry which transforms the minimal representation of $\EC$ to the representation implemented in the limit, which contains an infinite number of Kraus operators. Then after generalizing Theorem~\ref{thm1} to the case of general quantum instruments, Theorem~\ref{thm8}, we treated this same example as a (less than fully coarse-grained) quantum instrument, see Section~\ref{sec9}. For the latter case, it was shown \cite{WinterLeung} that this instrument cannot be exactly implemented by LOCC, but when considered as a channel, a proof of this statement is still lacking. Thus, it remains an open question whether LOCC$=\locc$ in the case of quantum channels. On the other hand, we have succeeded in answering this same question for the case of quantum measurements, which as far as we are aware, had not been previously answered. That is, we have given an example of a quantum measurement taken directly from the limiting Kraus representation for the above-mentioned $\EC$, which can be arbitrarily closely approximated, but not implemented exactly, by LOCC. This measurement consists of an infinite number of outcomes.
			
	We conjecture that LOCC is closed for measurements that are restricted to have a finite number of outcomes. It is certainly possible this is false, but our reasoning is as follows: That LOCC is not closed would mean that within $Z_\MC$, the zonotope associated with measurement $\MC$, any sequence of protocols whose limiting measurement is $\MC$ leads to a limiting path that is not piecewise local, since otherwise it would be LOCC. But as shown in our example of Section~\ref{sec5}, obtaining a path that isn't piecewise local appears to require that the sequence of approximating LOCC protocols involves paths that branch off at closer and closer intervals, leaving a continuum of branches in the limit. This continuum of branches implies a continuum of leaf nodes at the ends of those branches, and this suggests an infinite number of outcomes. However, it is possible to have multiple leaf nodes that are equal to each other, including possibly an infinite number of them. Therefore, it is not entirely clear whether or not the conjecture is correct. Of course, in our example we find that each leaf node is unique, and it is at least conceivable that this must be true for any case demonstrating this property of sequences of piecewise local paths having a limiting path that is not piecewise local.
	
	Finally, we would like to encourage further study of multipartite quantum channels, for which there appear to be relatively few examples of physical interest that have been studied in the literature. Hopefully, the results presented here will stimulate new interest in exploring such avenues of research.
	
\noindent\textit{Acknowledgments} --- We would like to thank Vlad Gheorghiu and Fei Shi for comments on an early version of this manuscript.	

	\appendix
	\section{Convergence of channels implies convergence of the associated zonoids}\label{secA}
	
	Suppose we have a sequence of LOCC protocols $\PC^\nu$, each implementing a quantum channel $\EC^\nu$, and $\lim_{\nu\to\infty}\EC^\nu=\EC$, where $\EC:\BC(\HC)\to\BC(\HC_o)$ and $\EC^\nu:\BC(\HC)\to\BC(\HC_o^\nu)$, having recognized that the input Hilbert space $\BC(\HC)$ must be the same for all these channels. (Recall that superscript $\nu$ denotes the position in the sequence of quantum channels and is not to be read as an exponent.) That is, for every $\delta>0$, there exists natural number $\mu$ such that for every $\nu>\mu$
	\begin{align}\label{eqnA1}
		\delta/16>\normm{\EC-\EC^\nu}_\diamond,
	\end{align}
	where $\normm{\cdot}_\diamond$ is the diamond norm, see \myeq{eqnA4} below. From the results of Ref.~\cite{continuityStinespring,InfoDisturbStinespring}, we have that% there exist Stinespring dilations $V,V^\nu$ for $\EC,\EC^\nu$, respectively, such that
	\begin{align}\label{eqnA2}
		\sqrt{\delta}/4>\sqrt{\normm{\EC-\EC^\nu}_\diamond}\ge\inf_{V,V^\nu}\normm{V-V^\nu}_\infty\ge\frac{\normm{\EC-\EC^\nu}_\diamond}{\sqrt{\normm{\EC}_\diamond}+\sqrt{\normm{\EC^\nu}_\diamond}}.
	\end{align}
	where the operator norm, $\normm{X}_\infty$, is equal to the largest singular value of $X$, and the infimum appearing here is over all Stinespring dilations \cite{Stinespring}, $V:\HC\to\HC_o\otimes\HC_e$ and $V^\nu:\HC\to\HC_o^\nu\otimes\HC_e^\nu$, of channels $\EC$ and $\EC^\nu$, respectively. Below, we will make use of channels, $\FC:\HC\to\HC_e$ complementary to $\EC$, and $\FC^\nu:\HC\to\HC_e^\nu$ complementary to $\EC^\nu$, and without loss of generality, we will consider minimal representations of $\EC$ $(\EC^\nu)$, so that the dimension of $\HC_e$ $(\HC_e^\nu)$ is equal to the Kraus rank $\kappa$ of $\EC$ ($\kappa^\nu$ of $\EC^\nu$). The dimensions of $\HC,\HC_o$ and $\HC_o^\nu$ are $d,d_o$ and $d_o^\nu$, respectively.% \cmmt{what norm is this on $V-V^\nu$? is it the operator trace norm? no, it is the infinity-norm}
	
	The trace norm of a quantum channel $T$ is defined as
	\begin{align}\label{eqnA3}
		\normm{T}_1=\max_{\normm{X}_1\le1}\normm{T(X)}_1,
	\end{align}
	where $\normm{X}_1=\Tr{\sqrt{X^\dag X}}$ is the trace norm of operator $X$. For $X\ge0$, we have that $\normm{X}_1=\Tr{X}$. With $\IC_\HC$ the identity channel on input Hilbert space $\HC$, recall the definition of the diamond norm as
	\begin{align}\label{eqnA4}
		\normm{T}_\diamond=\normm{T\otimes\IC_\HC}_1=\max_{\normm{X}_1\le1}\normm{(T\otimes\IC_\HC)X}_1.
	\end{align}
%	We can find a lower bound of this diamond norm over all quantum channels by choosing $X=\rho\otimes\sigma$ with $\rho,\sigma$ each positive semidefinite and having trace equal to unity, and then omitting the maximum in \myeq{eqnA4}. This tells us that for all quantum channels, $T$,
%	\begin{align}\label{eqnA5}
%		\normm{T}_\diamond\ge\normm{T(\rho)\otimes\sigma}_1=\Tr{T(\rho)}\Tr{\sigma}=1,
%	\end{align}
%	and we have used the fact that a quantum channel is trace-preserving. Using this result, we may rewrite \myeq{eqnA2} as \cite{InfoDisturbStinespring} \cmmt{this is wrong! Gives upper bound on expression on RHS of Eq. (A2), not lower bound. But just reference InfoDisturbStinespring and keep it simple by replacing RHS of Eq. (A2) with RHS of the following.}
	It is not difficult to show (by use of Eq.~($7$) in Ref.~\cite{ZyczChannelsEquidistant}, for example) that for any quantum channel $\EC$,
		\begin{align}\label{eqnA5}
				\normm{\EC}_\diamond=1.
		\end{align}
	Then, \myeq{eqnA2} becomes
	\begin{align}\label{eqnA6}
		\sqrt{\delta}/4>\sqrt{\normm{\EC-\EC^\nu}_\diamond}\ge\inf_{V,V^\nu}\normm{V-V^\nu}_\infty\ge\frac{1}{2}\normm{\EC-\EC^\nu}_\diamond.
	\end{align}
	From \myeq{eqnA1} we have an (arbitrarily small) upper bound proportional to $\delta$ on $\normm{\EC-\EC^\nu}_\diamond$, but we are here seeking to find a similar bound on the Hausdorff distance between corresponding zonotopes, $Z_\EC,Z_{\EC^\nu}$. The following arguments achieve that goal.
	
	While the infimum in \myeq{eqnA6} may not be achieved, it is always arbitrarily close to at least one point in the set over which it is taken. Therefore, for every $\delta>0$ there exists Stinespring dilations $\tilde V,\tilde V^\nu$ of $\EC,\EC^\nu$, respectively, such that $\sqrt{\delta}/2>\inf_{V,V^\nu}\normm{V-V^\nu}_\infty+\sqrt{\delta}/4\ge\normm{\tilde V-\tilde V^\nu}_\infty\ge\inf_{V,V^\nu}\normm{V-V^\nu}_\infty$. Therefore,
	\begin{align}\label{eqnA12}
		\sqrt{\delta}>2\normm{\tilde V-\tilde V^\nu}_\infty\ge\normm{\EC-\EC^\nu}_\diamond.
	\end{align}
	Since each of these isometries $\tilde V$ $(\tilde V^\nu)$ is a dilation not only of channel $\EC$ $(\EC^\nu)$ but also of a corresponding complementary channel $\FC$ $(\FC^\nu)$, it also holds that \cite{InfoDisturbStinespring}
	\begin{align}\label{eqnA7}
		\sqrt{\delta}>2\normm{\tilde V-\tilde V^\nu}_\infty\ge\normm{\FC-\FC^\nu}_\diamond.
	\end{align}
	Since the diamond norm is a maximum over inputs $X$, see \myeq{eqnA4}, we get a lower bound on this norm by evaluating $\FC-\FC^\nu$ acting on any density operator $X$, so choose $X=\ket{\Omega}\bra{\Omega}$, where $\ket{\Omega}=\sum_{i}\ket{i}\otimes\ket{i}/\sqrt{d}$ is a maximally entangled state on input Hilbert space $\HC\otimes\HC$. Then, we obtain
	\begin{align}\label{eqnA8}
		\sqrt{\delta}>2\normm{\tilde V-\tilde V^\nu}_\infty\ge\normm{\Phi-\Phi^\nu}_1,
	\end{align}
	where $\Phi=(I_\HC\otimes\FC)(\ket{\Omega}\bra{\Omega})$ is the Choi-Jamiolkowski operator \cite{Choi,Jamiolkowski} for channel $\FC$, and similarly for $\Phi^\nu$ and $\FC^\nu$.
	
	 Consider the two Kraus representations $\{K_m\}$ of $\EC$ with $d_e$ members, and $\{K_m^\nu\}$ of $\EC^\nu$ with $d_e^\nu$ members. Define $\tilde d_e=\max(d_e,d_e^\nu)$, expand the smaller of $\HC_e,\HC_e^\nu$ to dimension $\tilde d_e$, and add extra zero operators to the smaller of the two representations, so that these each have $\tilde d_e$ operators. For the smaller of these two Hilbert spaces, $\FC$ or $\FC^\nu$ still maps into the same unexpanded space, $\HC_e$ or $\HC_e^\nu$, which is now a subspace of the new, expanded space. This can be seen, in the case that $d_e<d_e^\nu$ for example, from the relation noted below \myeq{eqn11} of the main text, $\bra{l}K_m=\bra{m}Q_l$, implying that for those $m$ such that $K_m$ is the null operator, the added kets correspond to a subspace of $\HC_e$ which $\FC$ does \emph{not} map into. With these considerations, we have%\cmmt{still not sure the sum upper limits are correct???}
	\begin{align}\label{eqnA9}
		\Phi&=\frac{1}{d}\sum_{i,j=1}^{d}\sum_{l=1}^{d_o}\ket{i}\bra{j}\otimes Q_l\ket{i}\bra{j}Q_l^\dag\notag\\
		&=\frac{1}{d}\sum_{i,j=1}^{d}\sum_{m,m^\prime=1}^{\tilde d_e}\sum_{ l=1}^{d_o}\ket{i}\bra{j}\otimes\ket{m^\prime}\bra{m^\prime}Q_l\ket{i}\bra{j}Q_l^\dag\ket{m}\bra{m}\notag\\
		&=\frac{1}{d}\sum_{i,j=1}^{d}\sum_{m,m^\prime=1}^{\tilde d_e}\sum_{ l=1}^{d_o}\ket{i}\bra{j}\otimes\bra{l}K_{m^\prime}\ket{i}\bra{j}K_{m}^\dag\ket{l}\ket{m^\prime}\bra{m}\notag\\
		&=\frac{1}{d}\sum_{i,j=1}^{d}\sum_{m,m^\prime=1}^{\tilde d_e} \sum_{l=1}^{d_o}\ket{i}\bra{j}\otimes\bra{j}K_m^\dag\ket{l}\bra{l}K_{m^\prime}\ket{i}\ket{m^\prime}\bra{m}\notag\\
		&=\frac{1}{d}\sum_{i,j=1}^{d}\sum_{m,m^\prime=1}^{\tilde d_e}\bra{j}K_{m}^\dag K_{m^\prime}\ket{i}\ket{i}\bra{j}\otimes\ket{m^\prime}\bra{m},
	\end{align}
	and similarly for $\Phi^\nu$, the only difference being the replacement of $K_m$ by $K_m^\nu$. In going from the second to the third line of \myeq{eqnA9}, we have used the relationship, $\bra{m}Q_l=\bra{l}K_m$, see just below \myeq{eqn11} of the main text. Note that matrices $\Phi$ and $\Phi^\nu$ are of the same size, and we have
	\begin{align}\label{eqnA10}
		\normm{\Phi-\Phi^\nu}_1=\frac{1}{d}\normm{\sum_{i,j=1}^{d}\sum_{m,m^\prime=1}^{\tilde d_e}\bra{j}\left(K_{m}^\dag K_{m^\prime}-K_{m}^{\nu\dag} K_{m^\prime}^\nu\right)\ket{i}\ket{i}\bra{j}\otimes\ket{m^\prime}\bra{m}}_1.
	\end{align}
	
	Let $\hat K_n$ ($\hat K_n^\nu)$ be one of $\kappa$ $(\kappa^\nu)$ members of a minimal Kraus representation for $\EC$ $(\EC^\nu)$. Then, $K_m=\sum_nW_{mn}\hat K_n$ with $W$ a $\tilde d_e\times\kappa$ isometry so that $W^\dag W=I_\kappa$, and $K_m^\nu=\sum_nW_{mn}^\nu\hat K_n^\nu$ with $W^\nu$ a $\tilde d_e\times\kappa^\nu$ isometry so that $W^{\nu\dag }W^\nu=I_{\kappa^\nu}$. Recalling from Lemma~\ref{lem5} that $\kappa^\nu\ge\kappa$, add zero operators if needed to $\{\hat K_n\}$, so that this set also has $\kappa^\nu$ members, along with zero columns to $W$ so that $W^\dag W=P_\kappa$, a $\kappa^\nu\times\kappa^\nu$ projector onto a $\kappa$-dimensional subspace. Then, from Eqs.~(\ref{eqnA8}) and (\ref{eqnA10}), we can write
	\begin{align}\label{eqnA17}
		\sqrt{\delta}&>\frac{1}{d}\normm{\sum_{i,j=1}^{d}\sum_{m,m^\prime=1}^{\tilde d_e}\bra{j}\left(K_{m}^\dag K_{m^\prime}-K_{m}^{\nu\dag} K_{m^\prime}^\nu\right)\ket{i}\ket{i}\bra{j}\otimes\ket{m^\prime}\bra{m}}_1\notag\\
		&=\frac{1}{d}\left(\sum_{i,j=1}^d\sum_{m,m^\prime=1}^{\tilde d_e}\left\vert\bra{j}\left(K_{m}^\dag K_{m^\prime}-K_{m}^{\nu\dag} K_{m^\prime}^\nu\right)\ket{i}\right\vert^2\right)^{1/2}\notag\\
		&\ge\frac{1}{d}\sup_{0\le C\le I_{\tilde d_e}}\left(\sum_{i,j=1}^d\sum_{m,m^\prime=1}^{\tilde d_e} \left\vert C_{mm^\prime}\bra{j}\left(K_{m}^\dag K_{m^\prime}-K_{m}^{\nu\dag} K_{m^\prime}^\nu\right)\ket{i}\right\vert^2\right)^{1/2}\notag\\
		&\ge\frac{1}{d}\sup_{0\le C\le I_{\tilde d_e}}\left(\sum_{i,j=1}^d \norm{\sum_{m,m^\prime=1}^{\tilde d_e} C_{mm^\prime}\bra{j}\left(K_{m}^\dag K_{m^\prime}-K_{m}^{\nu\dag} K_{m^\prime}^\nu\right)\ket{i}}^2\right)^{1/2}\notag\\
		&=\frac{1}{d}\sup_{0\le C\le I_{\tilde d_e}}\inf_{0\le C^\nu\le I_{\tilde d_e}}\left(\sum_{i,j=1}^d \norm{\sum_{m,m^\prime=1}^{\tilde d_e}\sum_{n,n^\prime=1}^{\kappa^\nu}\bra{j}\left( W_{mn}^\ast C_{mm^\prime}W_{m^\prime n^\prime}\hat K_{n}^\dag \hat K_{n^\prime}-W_{mn}^{\nu\ast} C_{mm^\prime}^\nu W_{m^\prime n^\prime}^\nu\hat K_{n}^{\nu\dag}\hat K_{n^\prime}^\nu\right)\ket{i}}^2\right)^{1/2}\notag\\
		&=\frac{1}{d}\sup_{0\le\hat C\le P_{\kappa}}\inf_{0\le\hat C^\nu\le I_{\kappa^\nu}}\left(\sum_{i,j=1}^d \norm{\sum_{n,n^\prime=1}^{\kappa^\nu}\bra{j}\left( \hat C_{nn^\prime}\hat K_{n}^\dag \hat K_{n^\prime}- \hat C_{nn^\prime}^\nu\hat K_{n}^{\nu\dag}\hat K_{n^\prime}^\nu\right)\ket{i}}^2\right)^{1/2}\notag\\
		&=\frac{1}{d}\sup_{0\le\hat C\le P_{\kappa}}\inf_{0\le\hat C^\nu\le I_{\kappa^\nu}}\normm{\sum_{n,n^\prime=1}^{\kappa^\nu}\left(\hat C_{nn^\prime}\hat K_{n}^\dag \hat K_{n^\prime}- \hat C_{nn^\prime}^\nu\hat K_{n}^{\nu\dag}\hat K_{n^\prime}^\nu\right)}_1,
	\end{align}
	with $\hat C=W^\dag CW$ and $\hat C^\nu=W^{\nu\dag}C^\nu W^\nu$. To obtain the third line, we inserted $\norm{C_{mm^\prime}}\le1$, which will hold for all $m,m^\prime$ when $0\le C\le I_{\tilde d_e}$, and then we used the triangle inequality in the line after that. In entirely analogous fashion, one finds that
	\begin{align}\label{eqnA13}
		\sqrt{\delta}>\frac{1}{d}\sup_{0\le\hat C^\nu\le I_{\kappa^\nu}}\inf_{0\le\hat C\le P_{\kappa}}\normm{\sum_{n,n^\prime=1}^{\kappa^\nu}\left(\hat C_{nn^\prime}\hat K_{n}^\dag \hat K_{n^\prime}- \hat C_{nn^\prime}^\nu\hat K_{n}^{\nu\dag}\hat K_{n^\prime}^\nu\right)}_1.
	\end{align}
	Setting $\delta=(\epsilon/d)^2$, it immediately follows from Eqs.~\ref{eqnA1}, \ref{eqnA17} and \ref{eqnA13} that for every $\epsilon>0$ there exists $\mu$ such that for every $\nu>\mu$
	\begin{align}\label{eqnA14}
		\epsilon>d_H\left(Z_\EC,Z_{\EC^\nu}\right),
	\end{align}
	where $d_H(X,Y)=\max\{\sup_{x\in X}\inf_{y\in Y}\norm{x-y},\sup_{y\in Y}\inf_{x\in X}\norm{x-y}\}$ is the Hausdorff distance between sets $X,Y$, with $\norm{x-y}$ the distance between elements $x,y$, which for present purposes is the operator trace norm, $\normm{\cdot}_1$. 
	
	We therefore have that the sequence of zonoids, $Z_{\EC^\nu}$, associated with quantum channels $\EC^\nu$, converges to the zonoid $Z_\EC$, associated with the limiting quantum channel $\EC$. This completes the proof of Lemma~\ref{lem2}.
	
\section{Generalization of $Z_\EC$ for the case of quantum instruments}\label{secB}
	Assume that the sequence of instruments, $\EC^\nu$, has as its limit instrument $\EC=\{\JC_r\}$. To formalize the intuitive arguments leading up to the statement of Theorem~\ref{thm8} in Section~\ref{sec9}, let us consider $\EC_{qc}$ the \emph{quantum-classical} channel, $\rho\to\sum_r\JC_r(\rho)\otimes[r]_c$, uniquely associated with instrument $\EC$ \cite{WinterLeung}, and $\EC_{qc}^\nu$ associated with each $\EC^\nu$, as well. Then, $\lim_{\nu\to\infty}\EC_{qc}^\nu=\EC_{qc}$. Here, subscript $c$ on $[r]_c$ denotes an ancillary output containing the classical information as to which CP map, $\JC_r$, was obtained. As these are quantum channels, the analysis in Appendix~\ref{secA} holds, but let us take care in identifying the Kraus operators representing these channels. Recognizing that a set of Kraus operators for $\EC_{qc}$ is $K_m\otimes\ket{f(m)}_c$, where $f(m)=r$ for $K_m$ a Kraus operator representing $\JC_r$, then returning to \myeq{eqnA9} and noting that the first two lines there remain unchanged, we may follow through the same steps (but with minor re-arrangement of factors to make clear how the classical register is to be handled) to derive the Choi-Jamiolkowski operator for the channel complementary to $\EC_{qc}$. We find
\begin{align}
	\label{eqnB1}\Phi%&=\frac{1}{d}\sum_{ij}\sum_l\dyad{i}{j}\otimes Q_l\dyad{i}{j}Q_l^\dag\notag\\
	%&=\frac{1}{d}\sum_{ij}\sum_l\sum_{mm^\prime}\dyad{i}{j}\otimes\dyad{m^\prime}{m}\bra{m^\prime}Q_l\dyad{i}{j}Q_l^\dag\ket{m}\notag\\
	%&=\frac{1}{d}\sum_{ij}\sum_l\sum_{mm^\prime}\dyad{i}{j}\otimes\dyad{m^\prime}{m}\bra{j}Q_l^\dag\dyad{m}{m^\prime}Q_l\ket{i}\notag\\
	&=\frac{1}{d}\sum_{i,j=1}^d\sum_{m,m^\prime=1}^{\tilde d_e}\sum_{l=1}^{d_o}\dyad{i}{j}\otimes\dyad{m^\prime}{m}\bra{j}\left(K_m^\dag\otimes\bra{f(m)}_c\right)\left(\ket{l}\bra{l}\otimes I_c\right)\left(K_{m^\prime}\otimes\ket{f(m^\prime)}_c\right)\ket{i}\notag\\
	&=\frac{1}{d}\sum_{i,j=1}^d\sum_{m,m^\prime=1}^{\tilde d_e}\dyad{i}{j}\otimes\dyad{m}{m^\prime}\bra{j}K_m^\dag K_{m^\prime}\ket{i}\inpd{f(m)}{f(m^\prime)}_c\notag\\
	&=\frac{1}{d}\sum_{i,j=1}^d\sum_{m,m^\prime=1}^{\tilde d_e}\dyad{i}{j}\otimes\dyad{m}{m^\prime}\bra{j}K_m^\dag K_{m^\prime}\ket{i}\delta_{f(m),f(m^\prime)},
\end{align}
and similarly for $\Phi^\nu$ associated with each $\EC_{qc}^\nu$ but with a different coarse-graining map, $f^\nu(m)$. Due to the appearance of the Kronecker delta in the last line, this is a block-diagonal operator, each block associated with one of the CP maps $\JC_r$. Furthermore, each CP map has its own set of minimal Kraus operators, the collection of which (over all the CP maps in instrument $\EC$) we denote as $\{\hat K_n\}$. For the associated quantum-classical channel, these Kraus operators become $\hat K_n\otimes\ket{g(n)}_c$, where $g(n)$ is another coarse-graining map, possibly having a different domain than, but the same range as, $f(n)$. As before, $g(n)=r$ for those operators representing $\JC_r$. Then, isometric freedom of this channel implies
\begin{align}
	\label{eqnB2}K_m\otimes\ket{f(m)}_c&=\sum_nW_{mn}\hat K_n\otimes\ket{g(n)}_c\notag\\
		K_m^\nu\otimes\ket{f^\nu(m)}_c&=\sum_nW_{mn}^\nu\hat K_n^\nu\otimes\ket{g^\nu(n)}_c.
\end{align}
Clearly, since the kets appearing here represent classical states and are therefore mutually orthogonal (and we can always take them to be normalized by absorbing a factor into $K_m$ or $\hat K_n$, for example), there should be no contributions on the right-hand sides of these expressions except for those $n$ satisfying $g(n)=f(m)$ or $g^\nu(n)=f^\nu(m)$. That is to say, isometries $W~(W^\nu)$, introduced on the fifth line of \myeq{eqnA17}, are here block diagonal with blocks matching those appearing in $C~(C^\nu)$, also introduced in \myeq{eqnA17}. By arguments completely analogous to those in the proof of Lemma~\ref{lem5}, the blocks defined by $f^\nu(m)$ must \emph{contain} those blocks defined by $f(m)$, at least for large enough $\nu$. This means that a CP map of $\EC^\nu$ cannot have smaller Kraus rank than the corresponding CP map of $\EC$ (or possibly CP maps, plural, since the blocks defined by $f^\nu(m)$ may contain more than one of those defined by $f(m)$).

Thus, by continuing with the arguments following \myeq{eqnA9} in Appendix~\ref{secA}, one obtains the exact same result as previously obtained in Eqs.~(\ref{eqnA13}) and (\ref{eqnA14}), but with the additional condition that matrices $\hat C~(\hat C^\nu)$ appearing in the definition of $Z_\EC~(Z_{\EC^\nu})$ must be block diagonal. The remainder of the proof of Theorem~\ref{thm8} follows precisely the arguments in the proof of Theorem~\ref{thm1}.

\section{Proof of Lemma~\ref{lem4}}\label{secC}
	Define
	\begin{align}
		\label{eqn606}K_\alpha(s)=\sqrt{d\!\!\sqrt[\leftroot{-2}\uproot{2}P]{s}M(s)^{\otimes(\alpha-1)}\otimes[0]\otimes M(s)^{\otimes(P-\alpha)}},
	\end{align}
	where the $[0]$ appearing here is a local operator for party $\alpha$. Then, defining $\sigma=\sqrt[\leftroot{-2}\uproot{2}P]{s}-1$, we have that
	\begin{align}
		\label{eqn607}\int_{0}^{1}d\sigma K_\alpha(s)\dyad{i}{j}K_\alpha^\dag=\int_{0}^{1}d\sigma\sqrt{\sigma}^{(l_i+l_j-2)}\dyad{i}{j}=\frac{2}{l_i+l_j}\dyad{i}{j},
	\end{align}
	when the binary representation of $i$ and $j$ each have a $0$ in the $\alpha$th position, whereas this integral vanishes otherwise. Summing over all $\alpha$ gives an additional factor of $m_{ij}$. This accounts for all the Kraus operators except for the one equal to $[2^P-1]$, which gives $[2^P-1]$ when acting on itself, and zero otherwise. Therefore, adding all the terms together, we see that \myeq{eqn603} and \myeq{eqn605} in the main text are correct for channel $\EC_P$.
	
	For the approximating channel $\EC^\nu$, the Kraus operators are $K_M^\nu=\sqrt{R}^{\otimes\nu}$ at the end of the main branch, which asymptotically approaches $[2^P-1]$ in the limit $\nu\to\infty$ (equivalently, $\epsilon\to0$). In addition, we have $K_{\alpha n}^\nu=\sqrt{\epsilon(1-\epsilon)^{(n-1)}}\sqrt{R^n}^{\otimes(\alpha-1)}\otimes[0]\otimes\sqrt{R^n}^{\otimes(P-\alpha-1)}$, while all other Kraus operators are $\cal{O}(\epsilon)$ and will yield vanishing contributions in the limit. Now, $K_{\alpha n}^\nu\ket{i}\bra{j}K_{\alpha n}^{\nu\dag}=\epsilon(1-\epsilon)^{(n-1)}\left(1-\epsilon\right)^{n(l_i+l_j-2)/2}\dyad{i}{j}$ when $i,j$ each have a $0$ appearing at position $\alpha$ in their binary expansions. Summing over $\alpha$ gives a factor of $m_{ij}$, and then the sum over $n$ from $1$ to $\nu$ yields factor $2/(l_i+l_j)+\cal{O}(\epsilon)$. Therefore, the limiting channel also agrees with \myeq{eqn603} and \myeq{eqn605}, and the proof is complete.\endproof

%\bibliographystyle{quantum} % No DOI numbers if include this. Refs in alphabetical order without it.
%\bibliography{C:/Users/cohensm/Documents/Research/QRefs}	

\end{document}